\documentclass[12pt]{article}

\let\OLDthebibliography\thebibliography
\renewcommand\thebibliography[1]{
  \OLDthebibliography{#1}
  \setlength{\parskip}{0pt}
  \setlength{\itemsep}{7.7pt plus 0.3ex}
}

\usepackage[left=2.8cm,right=2.8cm,top=2.8cm,bottom=2.8cm]{geometry}

\usepackage[colorlinks=true,linkcolor=black,citecolor=black,urlcolor=blue,filecolor=black]{hyperref}

\usepackage{amsmath}
\usepackage{amsfonts}
\usepackage{amssymb}
\usepackage{stmaryrd}
\usepackage{mathtools}
\usepackage{mathrsfs}
\usepackage{color}
\usepackage{soul}
\usepackage[nosort]{cite}
\usepackage[normalem]{ulem}
 \csname
@addtoreset\endcsname{equation}{section}

\def\a{\alpha}
\def\b{\beta}

\def\d{\delta}

\def\e{\epsilon}

\def\th{\theta}

\def\k{\kappa}

\def\m{\mu}
\def\n{\nu}

\def\p{\pi}

\def\r{\rho}

\def\t{\tau}

\def\vf{\varphi}

\def\o{\omega}

\def\cA{{\cal A}}
\def\cB{{\cal B}}
\def\cC{{\cal C}}
\def\cD{{\cal D}}

\def\cF{{\cal F}}

\def\cH{{\cal H}}

\def\cK{{\cal K}}

\def\cN{{\cal N}}
\def\cO{{\cal O}}

\def\cR{{\cal R}}
\def\cS{{\cal S}}

\def\cW{{\cal W}}

\def\CC{\mathbb{C}}
\def\II{\mathbb{I}}
\def\NN{\mathbb{N}}
\def\RR{\mathbb{R}}

\def\ZZ{\mathbb{Z}}
\def\mg{\mathfrak{g}}

\def\be{\begin{equation}}
\def\ee{\end{equation}}
\def\bea{\begin{eqnarray}}
\def\eea{\end{eqnarray}}
\def\ba{\begin{array}}
\def\ea{\end{array}}

\def\nn{\nonumber}

\def\tr{\text{tr}}

\def\ww{\wedge}

\def\12{\frac{1}{2}}
\def\pr{\partial}

\def\dsl{\not {\! \pr}}

\def\psisl{\not {\!\! \psi}}

\def\ssl{\not {\! \cal S}}

\def\ssl{\not {\! \cal S}}

\begin{document}


\vspace{30pt}

\begin{center}


{\Large\sc Rotating Higher Spin Partition Functions\\[6pt]
and Extended BMS Symmetries}\\

---------------------------------------------------------------------------------------------------------


\vspace{25pt}
{\sc A.~Campoleoni${}^{\; a,}$\footnote{Postdoctoral Researcher of the Fund for Scientific Research-FNRS 
Belgium.}, H.A.~Gonzalez${}^{\; a}$, B.~Oblak${}^{\; a,b,}$\footnote{Research fellow of the Fund for 
Scientific Research-FNRS Belgium.} and M.~Riegler${}^{\; c}$}

\vspace{10pt}
{${}^a$\sl\small
Universit{\'e} Libre de Bruxelles\\
and International Solvay Institutes\\
ULB-Campus Plaine CP231\\
1050 Brussels,\ Belgium
\vspace{10pt}

${}^b$\sl\small 
DAMTP, Centre for Mathematical Sciences\\
University of Cambridge\\
Wilberforce Road\\
Cambridge CB3 0WA,\ United Kingdom
\vspace{10pt}

${}^c$\sl\small and
Institute for Theoretical Physics\\
Vienna University of Technology\\
Wiedner Hauptstrasse 8-10\\
1040 Vienna,\ Austria
\vspace{10pt}

{\it andrea.campoleoni@ulb.ac.be, hgonzale@ulb.ac.be,\\ boblak@ulb.ac.be, rieglerm@hep.itp.tuwien.ac.at} 
}

\vspace{50pt} {\sc\large Abstract} \end{center}

\noindent
We evaluate one-loop partition functions of higher-spin 
fields in thermal flat space with angular potentials; this computation is performed in arbitrary space-time 
dimension, and the result is a simple combination of Poincar\'e characters. We then focus on dimension 
three, showing that suitable products of one-loop partition functions coincide with vacuum characters of higher-spin asymptotic symmetry algebras at null infinity. These are extensions of the $\mathfrak{bms}_3$ algebra that emerges in pure gravity, and we propose a way to build their unitary representations and to compute the associated characters. We also extend our investigations to supergravity and to a class of gauge theories involving higher-spin fermionic fields.


\newpage


\tableofcontents


\section{Introduction}\label{sec:intro}

The structure of the known interacting field theories involving particles of spin greater than two depends 
significantly on the presence or absence of a cosmological constant. On (anti) de Sitter backgrounds of any 
dimension, Vasiliev's equations describe higher-spin gauge theories with an infinite tower of massless fields 
of increasing spin \cite{Vasiliev:2003ev}. These models display several unconventional features, mainly 
because interactions involve more than two derivatives. In flat space the situation is even subtler, since one 
loses the option to balance higher derivatives by inverse powers of the cosmological constant. As a result, 
interactions of massless fields in flat space are expected to be fraught with more severe non-localities than 
their (A)dS peers (see e.g.\ \cite{Bekaert:2010hw} for a review), and it is not yet clear if consistent 
interacting theories can be defined at all. These difficulties, however, are absent for massive fields, which 
bring in a dimensionful parameter that can play a role analogous to that of the cosmological constant. String 
field theories indeed involve infinitely many massive higher-spin fields and they can be defined on 
flat backgrounds.

Therefore, even if (A)dS backgrounds favour interactions of massless higher-spin particles, flat space is not 
completely ruled out by higher spins. Several indications also suggest that string models could actually be 
broken phases of a higher-spin gauge theory (see e.g.\ \cite{Sagnotti:2013bha} for a recent review). In order 
to clarify this issue one should understand if the striking differences in higher-spin theories with or 
without cosmological constant are really fundamental, or if they are induced by technical assumptions on the 
``allowed'' field theories. At present we indeed control higher-spin gauge theories only in a context in which 
we do not fully control String Theory. To make progress in this quest, it is important to develop tools to 
analyse the elusive higher-spin theories in flat space and to study the pathologies of the flat limit of the 
known interacting theories in (A)dS.\\

In this paper we consolidate one of these tools for flat space; we compute one-loop partition functions for 
higher-spin fields in $D$-dimensional Minkowski space at finite temperature and with non-vanishing angular 
potentials. Although they are determined by the free theory, one-loop partition functions often provide useful 
information on the consistency of a given spectrum for a possible interacting quantum field theory. This 
powerful feature has been extensively exploited for higher-spin gauge theories on AdS backgrounds: in $D=3$ 
the comparison between bulk and boundary partition functions 
\cite{Gaberdiel:2010ar,Gaberdiel:2011zw,Creutzig:2011fe} has been an important ingredient in defining the 
holographic correspondence between higher-spin gauge theories and minimal model CFTs \cite{Gaberdiel:2012uj}. 
In $D>3$ the analysis of one-loop partition functions of infinite sets of higher-spin fields provided the 
first quantum checks \cite{Giombi:2013fka,Giombi:2014iua,Giombi:2014yra,Beccaria:2014xda,Beccaria:2015vaa} of 
analogous AdS/CFT dualities \cite{Klebanov:2002ja}. More recently, holographic considerations driven by the 
structure of one-loop partition functions have also been used to conjecture the existence of consistent 
quantum higher-spin gauge theories in AdS with spectra that differ from those of Vasiliev's theories 
\cite{Tseytlin:2013jya,Beccaria:2014jxa,Basile:2014wua}.\\

In flat space this tool has been poorly employed in the higher-spin context; aside from computing one-loop partition functions for any $D$, here we also provide a first application of our results when $D=3$. This is a promising setup to explore relations between higher-spin theories in flat and (A)dS spaces because, in contrast with what happens when $D>3$, the limit of vanishing cosmological constant does not entail any subtlety. The main reason is that, in the absence of matter couplings, higher-spin gauge theories in AdS$_3$ are described by Chern-Simons actions \cite{Blencowe:1988gj} that can be cast in the form
\be \label{CSintro}
S = \frac{k_\mathfrak{g}}{16\p G} \int \tr \left( e \ww R + \frac{1}{3\ell^2}\, e\ww e\ww e \right) , \quad \text{with}\ R = d\o + \o \ww \o \, , 
\ee
where $e$ and $\o$ are one-forms that generalise the gravity vielbein and spin connection and that take values in a suitable gauge algebra (typically $\mathfrak{sl}(N,\RR)$ for theories involving fields of spin $2,3,\ldots,N$ --- see e.g.\ \cite{Campoleoni:2011tn} for a review). In \eqref{CSintro} $G$ denotes Newton's constant and $k_\mathfrak{g}$ is a factor that depends on the normalisation of the trace, while $\ell$ denotes the AdS radius. One can clearly consider the limit $\ell \to \infty$ in the action.\footnote{In the metric-like formulation of the dynamics \cite{Fronsdal:1978rb,Fang:1978wz}, on which we rely to compute partition functions, the limit is well defined because interactions involve at most two derivatives in $D=3$, so that no inverse powers of the cosmological constant enter the action \cite{Campoleoni:2014tfa}.} This simplification is related to the absence of local degrees of freedom in theories involving fields of spin $s \geq 2$. Another key feature of gravitational theories in three dimensions is the richness of their asymptotic symmetries in both AdS \cite{Brown:1986nw,Henneaux:2010xg,Campoleoni:2010zq,Gaberdiel:2011wb,Campoleoni:2011hg} and flat space \cite{Ashtekar:1996cd,Barnich:2006av,Barnich:2010eb,Afshar:2013vka,Gonzalez:2013oaa,Fuentealba:2015wza}. The combination of simplicity and powerful infinite-dimensional asymptotic symmetries makes these models important testing grounds for the holographic principle, in both its AdS/CFT realisation and its possible flat space counterpart. However, in spite of the straightforward way one can obtain interacting actions in flat space from the $\ell \to \infty$ limit of \eqref{CSintro}, higher-spin gauge theories in flat space are arguably less understood than those in AdS$_3$. An important reason is that their asymptotic symmetry algebras at null infinity are less familiar than those that emerge in AdS$_3$ at spatial infinity. The latter are typically $\cW_N$ algebras, which are well studied global symmetries in two-dimensional CFT \cite{Bouwknegt:1992wg}. In the following we propose a characterisation of the unitary representations of their flat space counterparts --- that we call ``flat $\cW_N$ algebras'' --- and we test our proposal by matching their vacuum characters with suitable products of partition functions of higher-spin fields. In this process we thus achieve two goals: on the one hand we improve the current understanding of the representation theory of flat $\cW_N$ algebras. On the other hand, in analogy with similar results in AdS$_3$  \cite{Giombi:2008vd,Gaberdiel:2010ar,Creutzig:2011fe}, we confirm that the asymptotic symmetries identified by the classical analysis of \cite{Afshar:2013vka,Gonzalez:2013oaa,Fuentealba:2015wza} are a robust feature of these models, that should persist also at the quantum level.\\

The paper is organised as follows: in sect.~\ref{sec:Z} we compute one-loop partition functions of higher-spin fields on Minkowski space of arbitrary dimensions $D$ with finite temperature $1/\beta$ and a maximal number of angular potentials $\vec{\th}$. We employ the heat kernel method of \cite{Giombi:2008vd}. As already shown for gravity in $D=3$ \cite{Barnich:2015mui}, these techniques are more tractable in flat space than in (A)dS, so that we do not have to resort to their successive refinements \cite{David:2009xg,Gopakumar:2011qs}. In sect.~\ref{subsec2.2} we study massive and massless bosonic fields of any (discrete) spin, whose partition functions are given in \eqref{ss6.5b} and (\ref{t6.5t},\,\ref{ss6q}). In sect.~\ref{subsec2.3} we then move to massive and massless fermionic fields of any spin, whose partition functions are given in \eqref{massivefermion}. In sect.~\ref{sec:poincare} we rewrite partition functions in terms of characters of the Poincar\'e group. For massive fields we obtain
\be
\label{Zintro}
Z_{M,s}[\beta,\vec\theta\,]
=
\exp\!\left[\,
\sum_{n\,=\,1}^{\infty}\frac{1}{n}\,\chi_{M,s}[n\vec{\theta},in\beta]
\,\right],
\ee
where $\chi_{M,s}$ is a character of a representation of the Poincar\'e group of mass $M$ and spin $s$. For massless fields this natural rewriting has to be amended when $D$ is odd, as one also has to introduce suitable angle-dependent coefficients.\\

In section \ref{sec:applications} we focus on $D =3$. In sect.~\ref{BMSPart} we begin by reviewing several 
aspects of the representation theory of the BMS$_3$ group --- i.e.\ of the group of asymptotic symmetries at 
null infinity of pure gravity in $D=3$ --- that are relevant in the following. We emphasise that its 
representations are induced representations classified by orbits of supermomenta, and we show how one can 
describe the representations of the corresponding $\mathfrak{bms}_3$ algebra in terms of induced modules. In 
sect.~\ref{sec:3DW} we move to higher spins, proposing to build unitary representations of flat $\cW_N$ 
algebras as Hilbert spaces of wavefunctions defined on coadjoint $\cW_N$ orbits of (higher-spin) supermomenta. 
We also compute vacuum characters and the characters of other illustrative representations. We then test our 
proposal by checking that vacuum characters of flat $\cW_N$ algebras --- which take the form
\be
\chi_{\text{vac}}[\th,\b]
=
e^{\frac{\beta}{8G}}
\prod_{s\,=\,2}^N
\left(\prod_{n\,=\,s}^{\infty}\frac{1}{|1-e^{in(\theta+i\epsilon)|^2}}\right),
\ee
where $\e$ is a regulator that ensures the convergence of the infinite product --- match the product of the partition functions of fields of spin $2,3,\ldots,N$ computed in sect.~\ref{subsec2.2}. We also take advantage of the description of these representations in terms of induced modules to make contact with previous proposals on the structure of representations of flat $\cW_N$ algebras. This allows us to explain how our representations evade some no-go arguments against the existence of unitary higher-spin gauge theories in three-dimensional flat space that appeared in the literature \cite{Grumiller:2014lna}. In sect.~\ref{sec:3DSugra} we include fermions: we discuss in detail the representation theory of the $\cN = 1$ extension of the BMS$_3$ group relevant for supergravity. We then extend our results to hypergravity theories describing the gravitational coupling of a massless field of spin $s + 1/2$. In both cases we also exhibit the matching between vacuum characters and the product of partition functions of fields of spin 2 and $s + 1/2$.\\

We close the paper with a discussion of possible extensions and applications of our work even beyond three dimensions (sect.~\ref{sec:future}), while two technical appendices fill the gap between the results of heat kernel computations and the rewriting of partitions functions in terms of Poincar\'e characters.

\section{Partition functions in flat space}\label{sec:Z}

We wish to study one-loop partition functions of higher-spin fields living in $D$-dimensional Minkowski space at 
finite temperature $1/\beta$, and with non-zero angular potentials. We will denote these potentials as 
$\vec\theta=(\theta_1,\ldots,\theta_r)$, where $r=\lfloor(D-1)/2\rfloor$ is the rank of $\text{SO}(D-1)$, that 
is, the maximal number of independent rotations in $(D-1)$ space dimensions. The computation 
involves a functional integral over fields living on a quotient of $\RR^D$, where the easiest way to 
incorporate one-loop effects is the heat kernel method. Accordingly, we will now briefly review this approach, before analysing separately bosons (sect.~\ref{subsec2.2}) and fermions (sect.~\ref{subsec2.3}). In sect.~\ref{sec:poincare} we then rewrite partition functions in terms of characters of the Poincar\'e group.

\subsection{Heat kernels and the method of images}

Our goal is to compute partition functions of the form
\be
\label{s2.5}
Z[\beta,\vec\theta\,]
=
\int\cD\phi\,e^{-S[\phi]}
\ee
where $\phi$ is some collection of fields (bosonic or fermionic) defined on a thermal quotient $\RR^D/\ZZ$ of flat Euclidean space, satisfying suitable (anti)periodicity conditions. (The explicit action of $\ZZ$ on $\RR^D$, with its dependence on $\beta$ and $\vec\theta$, will be displayed below --- see eq.~(\ref{2.6}).) The functional $S[\phi]$ is a Euclidean action for these fields. 
Expression (\ref{s2.5}) can be evaluated perturbatively around a saddle point $\phi_c$ of $S$, leading to the 
semi-classical (one-loop) result
\be
\label{ss2.5}
Z[\beta,\vec\theta\,]
\sim
e^{-S[\phi_c]}\left[\text{det}\!\left.\left(
\frac{\delta^2 S}{\delta\phi\delta\phi}
\right)\right|_{\phi_c}\right]^{\#}
\ee
where the exponent $\#$ depends on the nature of the fields that were integrated out. The quantity $\delta^2S/\delta\phi(x)\delta\phi(y)$ appearing in this expression is a differential operator acting on sections of a suitable vector bundle over $\RR^D/\ZZ$. The evaluation of the one-loop 
contribution to the partition function thus boils down to that of a functional determinant.\\

The heat kernel method is a neat way to compute such determinants; after gauge-fixing, they reduce to determinants of operators of the form $(-\Delta + M^2)$. In short (see e.g.~\cite{Vassilevich:2003xt,Giombi:2008vd} for details), it allows one to express $\text{det}(-\Delta+M^2)$ as 
an integral
\be
\label{2.5}
-\log\det(-\Delta+M^2)
=
\int_0^{\infty}
\frac{dt}{t}
\int d^Dx \,{\rm Tr}\left[K(t,x,x)\right],
\ee
up to an ultraviolet divergence that can be regulated with standard methods. Here $K(t,x,x')$ is a matrix-valued bitensor 
known as the {\it heat kernel} associated with $(-\Delta+M^2)$. It satisfies the heat equation
\be
\label{s4}
\frac{\partial}{\partial t}\,K(t,x,x')-(\Delta_x-M^2)\,K(t,x,x')=0 \,,
\ee
along with the initial condition
\begin{eqnarray}
\label{ss4}
K(t=0,x,x')=\delta^{(D)}(x-x')\,\mathbb{I} \, ,
\end{eqnarray}
with $\II$ the identity matrix having the same tensor structure as $K$ (here omitted for brevity).\\

Heat kernels are well suited for the computation of functional determinants on quotient spaces. Indeed, 
suppose $\Gamma$ is a discrete subgroup of the isometry group of $\RR^D$, acting freely on $\RR^D$. Introducing the 
equivalence relation $x\sim y$ if there exists a $\gamma\in\Gamma$ such
that $\gamma(x)=y$, we define the quotient manifold $\RR^D/\Gamma$ as the set of corresponding equivalence 
classes. Given a differential operator $\Delta$ on $\RR^D$, it naturally induces a 
differential operator on $\RR^D/\Gamma$, acting on fields that satisfy suitable (anti)periodicity conditions. 
Because the heat equation 
(\ref{s4}) is linear, the heat kernel on the quotient space can be obtained from the heat kernel on $\RR^D$ by the method of images:
\be
\label{t4}
K^{\mathbb{R}^D/\Gamma}(t,x,x')=\sum_{\gamma\,\in\,\Gamma }K (t,x,\gamma(x')) \, .
\ee
Here, abusing notation slightly, $x$ and $x'$ denote points both in $\RR^D$ and in its quotient. In writing (\ref{t4}) we are assuming, for simplicity, that 
the tensor structure of $K$ is trivial, but as soon as $K$ carries tensor or spinor indices 
(i.e.~whenever the fields under consideration have non-zero spin), the right-hand side involves Jacobians 
accounting for the non-trivial transformation law of $K$.\\

We will be concerned with thermal quantum field theories on rotating Minkowski space. This means we will define our fields on a quotient $\RR^D/\ZZ$ of Euclidean space, 
with the action of $\ZZ$ defined as follows. For odd $D$, we endow $\RR^D$ with Cartesian coordinates 
$(x_i,y_i)$ (where $i=1,\ldots,r$) and a Euclidean time coordinate $\tau$, so that an integer $n\in\ZZ$ acts on 
$\RR^D$ according to
\be
\label{2.6}
\gamma^n\begin{pmatrix} x_i \\ y_i \end{pmatrix}
=
\begin{pmatrix}
\cos(n\theta_i) & -\sin(n\theta_i)\\
\sin(n\theta_i) & \cos(n\theta_i)
\end{pmatrix} \cdot\begin{pmatrix} x_i \\ y_i \end{pmatrix}
\equiv
R(n\theta_i)\cdot\begin{pmatrix} x_i \\ y_i \end{pmatrix},
\quad
\gamma^n(\tau)
=
\tau+n\beta \, .
\ee
For 
even $D$ we simply add one more spatial coordinate $z$, invariant under $\ZZ$. In terms of the coordinates 
$\{x_1,y_1,\ldots,x_r,y_r,\tau\}$ (and possibly $z$ at the end of this list for even $D$), the Lorentz matrix 
implementing the rotation (\ref{2.6}) is the $n^{\text{th}}$ power of
\be
\label{2.7}
J
=
\begin{pmatrix}
 R(\th_1) & 0 & \cdots & 0 \\
0 &   \ddots & 0 & \vdots \\
\vdots & 0 &R(\th_r)   & 0 \\
0 & \cdots & 0 & 1
\end{pmatrix}
\quad\text{or}\quad
\begin{pmatrix}
 R(\th_1) & 0 & \cdots & 0 & 0\\
0 &   \ddots & 0 & \vdots & 0\\
\vdots & 0 &R(\th_r)   & 0 &0\\
0 & \cdots & 0 & 1&0\\
0 & \cdots & 0 & 0&1
\end{pmatrix}
\ee
for $D$ odd or $D$ even, respectively. Being isometries of flat space, these transformations are linear maps 
in Cartesian coordinates, and their $n^{\text{th}}$ power therefore coincides with the Jacobian matrix 
$\partial\gamma^n(x)^{\mu}/\partial x^{\nu}$ that will be needed later for the method of images. Throughout 
this paper we take all angles $\theta_1,\ldots,\theta_r$ to be non-vanishing. We now display the computation of 
one-loop partition functions on $\RR^D/\ZZ$, first for bosonic, then for fermionic higher-spin fields.

\subsection{Bosonic higher spins}\label{subsec2.2}

In this subsection we study the rotating one-loop partition function of a single bosonic field with spin $s$ and mass $M$ (including the massless case). For $M>0$ its Euclidean free action can be presented either (i) using a symmetric traceless field $\phi_{\mu_1\ldots\mu_s}$ of rank $s$ and a tower of auxiliary fields of ranks $s-2,s-3,\ldots, 0$  that do not display any gauge symmetry \cite{Singh:1974qz}, or (ii) using a set of doubly-traceless fields of ranks $s,s-1,\ldots,0$ subject to a gauge symmetry generated by traceless gauge parameters of ranks $s-1,s-2,\ldots,0$ \cite{Zinoviev:2001dt}. In the latter case, the quadratic action is given by the sum of Fronsdal actions \cite{Fronsdal:1978rb} for each of the involved fields, plus a set of cross-coupling terms with one derivative proportional to $M$ and a set of terms without derivatives proportional to $M^2$. In the massless limit, all couplings vanish and one can consider independently the Fronsdal action for the field of highest rank:
\be
\label{s3}
S[\phi_{\mu_1\ldots\mu_s}]
=
\frac{1}{2}\int d^Dx\,\phi^{\mu_1\ldots\mu_s}\left(
\cF_{\mu_1\ldots\mu_s}
- 
\frac{1}{2}\,\delta_{(\mu_1\mu_2}{\cF_{\mu_3\ldots\mu_s)\lambda}}^\lambda
\right),
\ee
where 
\be
\cF_{\mu_1 \cdots \mu_s}=\Box\,\phi_{\mu_1\ldots\mu_s}
-\pr_{(\mu_1|}\pr^{\lambda}\phi_{|\mu_2\ldots\mu_s)\lambda}
+\pr_{(\mu_1}\pr_{\mu_2}{\phi_{\mu_3\ldots\mu_s)\lambda}}^{\lambda} \, ,
\ee
and parentheses denote the symmetrisation of the indices they enclose, with the minimum number of terms needed and without any overall factor. In the alternative formulation of the dynamics where no gauge symmetry is present \cite{Singh:1974qz}, in the massless limit all auxiliary fields except the one of rank $s-2$ decouple. The remaining fields can be combined into a doubly traceless field whose action is given again by \eqref{s3} \cite{Fronsdal:1978rb}. For further details we refer e.g.\ to \cite{Rahman:2013sta}.

Note that in all space-time 
dimensions other than three, the vacuum saddle point of the action (\ref{s3}) (or of its massive counterpart) is 
the trivial field configuration $\phi_{\mu_1\ldots\mu_s}=0$. Accordingly, the whole partition function 
(\ref{ss2.5}) is captured by its one-loop piece. In $D=3$, the presence of a mass gap makes this situation 
slightly different; we will return to this issue at the end of this subsection.

\subsubsection*{Massive case}

Applying e.g.\ the techniques of \cite{Gaberdiel:2010ar} to the presentation of the Euclidean action of a massive field of spin $s$ of \cite{Zinoviev:2001dt}, one finds that the partition function is given by
\be
\label{ss5b}
\log Z
=
-\,\frac{1}{2}\log \det (-\Delta^{(s)}+M^2)+\frac{1}{2}\log \det (-\Delta^{(s-1)}+M^2) \, ,
\ee
where $\Delta^{(s)}$ is the Laplacian $\partial_{\mu}\partial^{\mu}$ acting on periodic,\footnote{More precisely, the field at Euclidean time $\tau+\beta$ is rotated by $\vec\theta$ with respect to the field at time $\tau$.} symmetric, traceless 
tensor fields with $s$ indices on $\RR^D/\ZZ$. We will denote the 
heat kernel associated with $(-\Delta^{(s)}+M^2)$ on $\RR^D$ as $K_{\mu_s,\nu_s}(t,x,x')$, where $\mu_s$ and 
$\nu_s$ are shorthands that denote sets of $s$ symmetrised indices.
The heat 
equation (\ref{s4}) and initial condition (\ref{ss4}) for $K_{\mu_s,\nu_s}(t,x,x')$ then read
\be
\label{2.8}
(\Delta^{(s)}-M^2-\pr_t)K_{\mu_s,\,\nu_{s}}=0 \, , \quad K_{\mu_s,\,\nu_s}(t=0,x,x')=\II_{\mu_s,\,\nu_s} 
\delta^{(D)}(x-x') \, ,
\ee
where $\II_{\mu_s,\nu_s}$ is an identity matrix with the same tensor structure as $K_{\mu_s,\nu_s}$. A set of repeated covariant or contravariant indices also denotes a set of indices that are symmetrised with the minimum number of terms required and without multiplicative factors, while contractions involve as usual a covariant and a contravariant index. The tracelessness condition on the heat kernel amounts e.g.\ to
\be
\label{2.9}
\delta^{\mu \mu} {K}_{\mu_{s},\,\nu_s}= \delta^{\nu \nu} K_{\mu_s,\,\nu_{s}}=0 \, .
\ee
The solution of (\ref{2.8}) fulfilling this condition is
\be
\label{s6}
K_{\mu_s,\,\nu_s}(t,x,x')
=
\frac{1}{(4\pi t)^{D/2}}\,
e^{-M^2 t-\frac{1}{4t}|x-x'|^2}\,
\II_{\mu_s,\,\nu_s}\,,
\ee
with
\be
\II_{\mu_s,\,\nu_s}
=
\sum^{\lfloor\frac{s}{2}\rfloor}_{n=0}
\frac{(-1)^n 2^n n!\,[D+2(s-n-2)]!!}
{s!\,[D+2(s-2)]!!}\, \d_{\m\m}^n \d_{\m\n}^{s-2n} \d_{\n\n}^n\,.
\ee
Note that the dependence of this heat kernel on the space-time points $x$, $x'$ and on Schwinger proper time 
$t$ is that of a scalar heat kernel, and completely factorises from its spin/index structure which is wholly 
accounted for by the matrix $\II$. This simplification is the main reason why explicit heat kernel 
computations are more tractable in flat space than in AdS or dS.\\

To determine the heat kernel associated with the operator $(-\Delta^{(s)}+M^2)$ on $\RR^D/\ZZ$, we use the 
method of images (\ref{t4}), taking care of the non-trivial index structure. Denoting the matrix (\ref{2.7}) 
by ${J_{\alpha}}^{\beta}$ (it is the Jacobian of the transformation $x\mapsto \gamma(x)$), the spin-$s$ heat 
kernel on $\RR^D/\ZZ$ is
\be
K^{\mathbb{R}^D/\mathbb{Z}}_{\mu_s,\,\nu_s}(t,x,x')
=
\sum_{n\,\in\,\ZZ}{(J^n)_{\a}}^{\b}\ldots{(J^n)_{\a}}^{\b}
K_{\mu_s,\,\beta_s}(t,x,\gamma^n(x')) \, ,
\ee
where we recall again that repeated covariant or contravariant indices are meant to be symmetrised with the minimum number of terms required and without multiplicative factors, while repeating a covariant index in a contravariant position denotes a contraction.
Accordingly, formula (\ref{2.5}) gives the determinant of $(-\Delta^{(s)}+M^2)$ on $\RR^D/\ZZ$ as
\be
\begin{split}
& -\log\det(-\Delta^{(s)}+M^2) =
\int_0^{+\infty} \frac{dt}{t}
\int_{\RR^D/\ZZ} d^Dx\,
(\delta^{\mu \alpha})^s\,
K^{\mathbb{R}^D/\mathbb{Z}}_{\mu_s,\,\alpha_s} (t,x,x) \\[3pt]
& =
\sum_{n\,\in\,\ZZ}
(J^n)^{\mu\beta}\cdots(J^n)^{\mu\beta}\,
\II_{\mu_s,\beta_s}
\int_0^{+\infty} \frac{dt}{t}
\int_{\RR^D/\ZZ} d^Dx\,
\frac{1}{(4\pi t)^{D/2}}\,
e^{-M^2 t-\frac{1}{4t}|x-\gamma^n(x)|^2}.\quad\quad\label{s5t}
\end{split}
\ee
\label{t5t}In this series the term $n=0$ contains both an ultraviolet divergence (due to the singular 
behaviour of the integrand as $t\rightarrow 0$) and an infrared one (due to the integral of a constant over 
$\RR^D/\ZZ$), proportional to the product $\beta V$ with $V$ the spatial volume of the system. This 
divergence is a quantum contribution to the vacuum energy, which we ignore from now on. The only non-trivial one-loop contribution we must take into account then comes 
from the terms $n\neq0$ in (\ref{s5t}). Using
\be
|x-\gamma^n(x)|^2
=
n^2\beta^2+\sum_{i\,=\,1}^r4\sin^2(n\theta_i/2)(x_i^2+y_i^2)
\ee
in terms of the coordinates introduced around (\ref{2.6}), the integrals over $t$ and $x$ give
\be
\label{ss5t}
-\log \det(-\Delta^{(s)}+M^2)
=
\sum_{n\,\in\,\ZZ^*}
\frac{1}{|n|}
\frac{\chi_s[n\vec\theta,\vec{\e}\,]}{\prod\limits_{j=1}^r|1-e^{in(\theta_j+i\epsilon_j)}|^2}
\times
\begin{cases}
e^{-|n|\beta M}  &\mbox{if } D \ { \rm odd},\\[3pt]
\frac{M\Delta z}{\pi}K_1(|n|\beta M)& \mbox{if } D \  {\rm even},
\end{cases}
\ee
where $K_1$ is the first modified Bessel function of the second 
kind, and $\Delta z\equiv\int_{-\infty}^{+\infty}dz$ is an infrared divergence that arises in even dimensions 
because the $z$ axis is left fixed by the rotation (\ref{2.7}). Following \cite{Barnich:2015mui} we have 
added small imaginary parts to the angles $\theta_j$ to make the series convergent. Similarly 
\be
\label{s5.5t}
\chi_s[n\vec\theta,\vec{\e}\,]
\equiv
(J^n)^{\mu\beta}\ldots(J^n)^{\mu\beta}\,
\II_{\mu_s,\,\beta_s} \equiv \left[(J^n)^{\mu\beta}\right]^s \II_{\mu_s,\,\beta_s}
\ee
is the full mixed trace of $\II_{\mu_s,\nu_s}$, with the understanding that $\theta_j$ is replaced by $\theta_j\pm i\epsilon_j$ in all positive powers of $e^{\pm i\theta_j}$. For odd $D$, the result of this regularisation agrees with the flat limit of the AdS one-loop determinant, in which case the parameters $\epsilon_j=\beta/\ell$ are remnants of the inverse temperature (with $\ell$ the AdS radius).\footnote{According to \cite{Prohazka2015}, a different regularisation may produce a completely different, albeit finite, result.} For even $D$, the flat limit of the AdS result contains an infrared divergence; it is not obvious to us how this divergence can be regularised so as to reproduce the combination $\Delta z\cdot K_1$ of (\ref{ss5t}), but apart from this subtlety, the other terms of the expression indeed coincide with the flat limit of their AdS counterparts. From now on we will often omit displaying the $\epsilon$-regularisation explicitly, keeping it only in the final results.\\

Expression (\ref{ss5t}) is a higher-dimensional, 
higher-spin extension of the result derived previously for spin two in three dimensions in 
\cite{Barnich:2015mui}. In particular, the divergence as $\epsilon_j\rightarrow0$ is the same as in three 
dimensions. The new ingredient is the angle-dependent trace (\ref{s5.5t}); in appendices \ref{AppA1} and \ref{AppA2} we show that it is the character of an irreducible, unitary representation of 
$\text{SO}(D)$ with highest weight $\lambda_s \equiv (s,0,\ldots,0)$. More 
precisely, let $H_i$ denote the generator of rotations in the plane $(x_i,y_i)$, in the coordinates 
defined around (\ref{2.6}). Then the Cartan subalgebra $\mathfrak{h}$ of $\mathfrak{so}(D)$ is generated by 
$H_1,\ldots,H_r$, plus, if $D$ is even, a generator of rotations in the plane $(\tau,z)$. Denoting the dual 
basis of $\mathfrak{h}^*$ by $L_1,\ldots,L_r$ (plus possibly $L_{r+1}$ if $D$ is even), we can consider the 
weight $\lambda_s=sL_1$ whose only non-zero component (in the basis of $L_i$'s) is the first 
one. The 
character of the corresponding highest-weight representation of $\mathfrak{so}(D)$ coincides with expression (\ref{s6.5}):
\be
\label{s6b}
\chi_s[n\vec\theta\,]
=
\chi_{\lambda_s}[n\theta_1,\ldots,n\theta_r]
\quad\text{or}\quad
\chi_{\lambda_s}[n\theta_1,\ldots,n\theta_r,0] \, ,
\ee
for $D$ odd or even, respectively.\\

We can finally display an explicit formula for the one-loop partition function (\ref{ss5b}). Using expression 
(\ref{ss5t}) for the one-loop determinant together with property (\ref{s6b}), 
we find
\be
Z[\beta,\vec\theta\,]= 
\exp\!
\left[
\sum_{n=1}^{\infty}
\frac{n^{-1}}{\prod\limits_{j=1}^r|1-e^{in\theta_j}|^2} \times
\begin{cases}
\left(
\chi_{\lambda_s}^{\text{SO}(D)}[n\vec\theta\,]
-
\chi_{\lambda_{s-1}}^{\text{SO}(D)}[n\vec\theta\,]
\right)
e^{-n\beta M}\\[10pt]
\left(
\chi_{\lambda_s}^{\text{SO}(D)}[n\vec\theta,0]
-
\chi_{\lambda_{s-1}}^{\text{SO}(D)}[n\vec\theta,0]
\right)\!
\frac{M\Delta z}{\pi}K_1(n\beta M)
\end{cases}
\!\!\!\!\!\!\right],
\label{s6.5b}
\ee
where the upper (resp.\ lower) line corresponds to the case where $D$ is odd (resp.\ even). Remarkably, the 
differences of $\text{SO}(D)$ characters appearing here can be simplified: according to eqs.~(\ref{app:SO(D)DifferenceProofRelations1}) and (\ref{s37.5}), the difference of two $\text{SO}(D)$ characters with weights $(s,0,\ldots,0)$ and $(s-1,0,\ldots,0)$ 
is a (sum of) character(s) of $\text{SO}(D-1)$:
\be
\label{t6.5b}
\begin{rcases*}
\chi_{\lambda_s}^{\text{SO}(D)}[\vec\theta\,]
-
\chi_{\lambda_{s-1}}^{\text{SO}(D)}[\vec\theta\,] & (D\text{ odd})\\[5pt]
\chi_{\lambda_s}^{\text{SO}(D)}[\vec\theta,0]
-
\chi_{\lambda_{s-1}}^{\text{SO}(D)}[\vec\theta,0] & (D\text{ even})
\end{rcases*}
=
\chi_{\lambda_s}^{\text{SO}(D-1)}[\vec\theta\,] \, .
\ee
Since $r=\lfloor(D-1)/2\rfloor$ is the rank of $\text{SO}(D-1)$, the right-hand side of this equality makes 
sense regardless of the parity of $D$, and the partition function (\ref{s6.5b}) boils down to
\be
\label{ss6.5b}
Z[\beta,\vec\theta\,]
=
\exp\!
\left[\,
\sum_{n=1}^{\infty}
\frac{1}{n}
\frac{\chi_{\lambda_s}^{\text{SO}(D-1)}[n\vec{\theta},\vec{\e}\,]}
{\prod\limits_{j=1}^r|1-e^{in(\theta_j+i\epsilon_j)}|^2}
\times
\begin{cases}
e^{-n\beta M} & (D\text{ odd})\\[5pt]
\frac{M\Delta z}{\pi}K_1(n\beta M) & (D\text{ even})
\end{cases}
\,\right].
\ee
In sect.~\ref{sec:poincare} we will show that the function 
of $n\vec\theta$ and $n\beta$ appearing here in the sum over $n$ is in fact the character of an irreducible, 
unitary representation of the Poincar\'e group with mass $M$ and spin $s$ (see eq.~(\ref{Zexp})). A 
similar result holds in Anti-De Sitter space \cite{Gibbons:2006ij,Gopakumar:2011qs,Beccaria:2015vaa}.

\subsubsection*{Massless case}

We now turn to the one-loop partition function associated with the Fronsdal action (\ref{s3}). The gauge symmetry forces one to fix a gauge and introduce ghost fields \cite{Gaberdiel:2010ar}, and leads to the following expression for the one-loop term of the partition function:
\be
\label{2.4bos}
\log Z
=
-\frac{1}{2}\log \det (-\Delta^{(s)})+\log\det(-\Delta^{(s-1)})-\frac{1}{2}\log\det(-\Delta^{(s-2)}) \, .
\ee
As before, $\Delta^{(s)}$ is the Laplacian on $\RR^D/\ZZ$ acting on periodic, traceless, symmetric fields 
with $s$ indices. The functional determinants can be evaluated exactly as in the massive case, setting $M=0$. 
In particular, using $\lim_{x\rightarrow0}xK_1(x)=1$, the massless version of the functional determinant 
(\ref{ss5t}) is
\be
\label{s6.5t}
-\log \det(-\Delta^{(s)})
=
\sum_{n\,\in\,\ZZ^*}
\frac{1}{|n|}
\frac{\chi_s[n\vec\theta,\vec{\e}\,]}{\prod\limits_{j=1}^r|1-e^{in(\theta_j+i\epsilon_j)}|^2}
\times
\begin{cases}
1  &\mbox{if } D \ { \rm odd},\\
\frac{\Delta z}{\pi|n|\beta}&\mbox{if } D \  {\rm even},
\end{cases}
\ee
which has been regularised as discussed in the massive case. The matching (\ref{s6b}) between $\chi_s$ and a character of $\text{SO}(D)$ remains valid, but a sharp 
difference arises upon including all three functional determinants in 
(\ref{2.4bos}): the combination of $\chi_s$'s is
\be
\label{ss6.5t}
\chi_s[n\vec\theta\,]-2\chi_{s-1}[n\vec\theta\,]+\chi_{s-2}[n\vec\theta\,]
\stackrel{\text{(\ref{s6b})-(\ref{t6.5b})}}{=}
\chi_{\lambda_s}[n\vec\theta\,]
-
\chi_{\lambda_{s-1}}[n\vec\theta\,] \,.
\ee
It is tempting to use (\ref{t6.5b}) once more to rewrite this as a character of $\text{SO}(D-2)$, and indeed 
this is exactly what happens for even $D$ because in this case the rank of $\text{SO}(D-1)$ equals that of 
$\text{SO}(D-2)$:
\be
\label{t6.5t}
Z[\beta,\vec\theta]
=
\exp
\left[
\sum_{n\,=\,1}^{\infty}
\frac{1}{n}
\frac{\chi_{\lambda_s}^{\text{SO}(D-2)}[n\vec{\theta},\vec{\e}\,]}
{\prod\limits_{j=1}^r|1-e^{in(\theta_j+i\epsilon_j)}|^2}
\frac{\Delta z}{\pi n\beta}
\right]
\quad(\text{even }D).
\ee
If $D$ is {\it odd}, however, in going from $\text{SO}(D-1)$ to $\text{SO}(D-2)$, the rank 
decreases by one unit: expression (\ref{ss6.5t}) contains one angle too much to be a character of 
$\text{SO}(D-2)$. As we show in appendix \ref{AppA3}, one can nevertheless write the difference (\ref{ss6.5t})  as a sum of $\text{SO}(D-2)$ characters with angle-dependent coefficients (see 
eq.~(\ref{app:SO(D)DifferenceProofRelations2})). Namely, let us define 
\be
\label{s6q}
\cA_k^r(\vec\theta)
\equiv
\frac{|\cos((r-i)\theta_j)|_{\theta_k=0}}{|\cos((r-i)\theta_j)|}\,,
\quad k=1,\ldots,r,
\ee
where $|A_{ij}|$ denotes the determinant of an $r\times r$ matrix. Then the rotating one-loop partition 
function for a massless field with spin $s$ reads
\be
\label{ss6q}
Z[\beta,\vec\theta]
=
e^{-S^{(0)}}
\exp
\left[
\sum_{n\,=\,1}^{\infty}
\frac{1}{n}
\frac{\sum\limits_{k=1}^r\cA_k^r(n\vec\theta, \vec{\e})\,
\chi_{\lambda_s}^{\text{SO}(D-2)}[n\theta_1,\ldots,\widehat{n\theta_k},\ldots,n\theta_r, \vec{\e}\,]}
{\prod\limits_{j=1}^r|1-e^{in(\theta_j+i\epsilon_j)}|^2}
\right]
\quad(\text{odd }D),
\ee
where the hat on top of an argument denotes omission. We have also included a spin-dependent classical 
action $S^{(0)}$, whose value is a matter of normalisation and is generally taken to vanish, except in $D=3$. In the latter case, $S^{(0)}=-\beta/8G$ for $s=2$ (where $G$ is 
Newton's constant) ensures invariance of the on-shell action under modular transformations of the 
vacuum\cite{Barnich:2015mui,Bagchi:2013lma}, in analogy with the similar choice in AdS$_3$ 
\cite{Giombi:2008vd}. On the other hand, $S^{(0)}=0$ for all other spins because their vacuum expectation 
values are assumed to vanish. For $D=3$ the partition function (\ref{ss6q}) 
can thus be written as
\be
\label{ss6.5q}
Z[\beta,\vec\theta]
=
e^{\delta_{s,2} \frac{\beta c_2}{24}}
\prod_{n\,=\,s}^{\infty}\frac{1}{|1-e^{in(\theta+i\epsilon)}|^2}\,,
\quad
c_2=3/G,
\ee
and is the flat limit of the analogous higher-spin partition function in AdS$_3$ 
\cite{Gaberdiel:2010ar}. We will return to this formula in sect.~\ref{sec:3DW}.\\

The massless partition functions (\ref{t6.5t}) and (\ref{ss6q}) are related to the massless limit of 
(\ref{ss6.5b}). Indeed, as we show in appendix \ref{AppA4}, it turns out that
\be
\label{t6q}
\chi_{\lambda_s}^{\text{SO}(D-1)}[\vec\theta\,]
=
\sum_{j=0}^s
\begin{cases}
\sum_{k=1}^r\cA^r_k(\vec\theta)\,
\chi_{\lambda_j}^{\text{SO}(D-2)}[\theta_1,\ldots,\widehat{\theta_k},\ldots,\theta_r]
& \text{ for odd }D,
\\[8pt]
\chi_{\lambda_j}^{\text{SO}(D-2)}[\vec \theta]
& \text{ for even }D.
\end{cases}
\ee
Accordingly, the massless limit of a massive partition function with spin $s$ is a product of massless partition functions with spins ranging from $0$ to $s$,
\be
\label{s6.5q}
\lim_{M\rightarrow0}Z_{M,s}
=
\prod_{j\,=\,0}^sZ_{\text{massless},j},
\ee
consistently with the structure of the action \cite{Zinoviev:2001dt}. This result stresses again the role of the functions 
$\cA_k^r(\vec\theta\,)$ defined in \eqref{s6q}: when the dimension of the space-time is odd one needs angular dependent coefficients because the rank of the little group of massless particles is smaller than the maximum number of angular velocities, so that a single $SO(D-2)$ character cannot account for all of them.

\subsection{Fermionic higher spins}\label{subsec2.3}

We now turn to the fermionic analogue of the analysis of the previous subsection. The Euclidean action for a field with spin 
$s+1/2$ (where $s$ is a non-negative integer) and mass $M > 0$ can be given either (i) using a symmetric, $\gamma$-traceless spinor field with $s$ 
space-time indices and a set of auxiliary fields with no gauge symmetry \cite{Singh:1974rc} or (ii) using a set of symmetric spinor fields with $s,s-1,\ldots,0$ space-time indices and vanishing triple $\gamma$-trace, subject to a gauge symmetry generated by $\gamma$-traceless parameters with $s-1,\ldots,0$ space-time indices \cite{Metsaev:2006zy}. In the latter case, the action is given again by a sum of actions for massless fields of each of the involved spins, plus a set of cross-coupling terms proportional to the mass. In the limit $M \to 0$ the quadratic couplings vanish and one is left with a sum of decoupled Fang-Fronsdal actions \cite{Fang:1978wz}
\be
\label{ss8.5}
S[\psi,\bar\psi]
=
\frac{1}{2}\int d^Dx\,
\bar{\psi}^{\mu_1\ldots\mu_s}
\left(
\cS_{\mu_1\ldots\mu_s}-\frac{1}{2}\,\gamma_{(\mu_1}\!\!\ssl_{\mu_2 \cdots \mu_s)}
-
\frac{1}{2}\,\delta_{(\mu_1\mu_2 }{\cS_{\mu_3 \cdots \mu_s)\lambda}}^\lambda + {\rm h.c.}
\right),
\ee
where
\be
\cS_{\mu_1\ldots\mu_s}=i\left(\dsl\,\psi_{\mu_1\ldots\mu_s}-\pr_{(\mu_1}\!\!\psisl{}_{\mu_2\ldots\mu_s)}\right) ,
\ee
and one can consider only the field of highest rank/spin.\\

To compute the partition function for $\psi$, $\bar\psi$ we need to evaluate a path integral (\ref{s2.5}) 
with the integration measure $\cD\psi\cD\bar\psi$ and $S$ the action (\ref{ss8.5}) or its massive analogue. 
The fermionic fields live on $\RR^D/\ZZ$ as defined by the group action (\ref{2.6}), but in contrast to 
bosons, they satisfy {\it anti}periodic boundary conditions along the thermal cycle. For a massive field, 
one thus finds that the partition function is given by
\be
\label{2.18}
\log Z
=
\frac{1}{2}
\log\det(-\Delta^{(s+1/2)}+M^2)
-\frac{1}{2}\log\det(-\Delta^{(s-1/2)}+M^2)\,,
\ee
where $\Delta^{(s+1/2)}$ is the Laplacian acting on antiperiodic, symmetric, $\gamma$-traceless spinor fields 
with $s$ indices on $\RR^D/\ZZ$. For massless fields, the gauge symmetry enhancement requires gauge-fixing 
and ghosts, leading to \cite{Creutzig:2011fe}
\be
\label{2.4fer}
\log Z
=
\frac{1}{2}
\log\det(-\Delta^{(s+1/2)})
-
\log\det(-\Delta^{(s-1/2)})
+\frac{1}{2}
\log\det(-\Delta^{(s-3/2)})\,.
\ee
To evaluate the necessary functional determinants, we will rely once more on heat kernels and the method of 
images.\\

The heat kernel ${\cK^{AB}}_{\mu_s,\nu_s}$ associated with the operator $(-\Delta^{(s+1/2)}+M^2)$ on $\RR^D$ is 
the unique solution of the heat equation
\be
\label{2.14}
({\Delta_{(s+1/2)}}-M^2-\pr_t){\cK^{AB}}_{\mu_s,\nu_s}=0\,,
\quad
{\cK^{AB}}_{\mu_s,\,\nu_s}(t=0,x,x')= 
\II^{(F)}_{\mu_s,\,\nu_s}{\bf 1}^{AB}\delta^{(D)}(x-x')\,.
\ee
Here ${\cK^{AB}}_{\mu_s,\nu_s}$ is a bispinor in the indices $A$ and $B$, and a symmetric bitensor 
in the indices $\mu_s$ and $\nu_s$. (We use again the shorthand $\mu_s$ to denote a set of $s$ symmetrised 
indices.) It is also $\gamma$-traceless in the sense that 
\be
\label{2.15}
\gamma^{\mu}{\cK}_{\mu_{s},\,\nu_s}={\cK}_{\mu_s,\,\nu_{s}}\gamma^{\nu}=0 \, .
\ee
The 
solution of (\ref{2.14}) satisfying this requirement is
\be
\label{s9}
{\cK}_{\mu_s,\,\nu_s}(t,x,x')
=
\frac{1}{(4\pi t)^{D/2}}
e^{-M^2 t-\frac{1}{4t}|x-x'|^2}\,\II^{(F)}_{\mu_s,\,\nu_s}\,,
\ee
where $\II^{(F)}_{\mu_s,\nu_s}$ is the following bisymmetric, $\gamma$-traceless tensor:
\be
\II^{(F)}_{\mu_s,\,\nu_s}
=
\sum_{k\,=\,0}^{\lfloor\frac{s}{2}\rfloor}
\frac{(-1)^k 2^k k!\, [D+2(s-k-1)]!!}{s!\,[D+2(k-1)]!!}
\left(
\delta^k_{\mu \mu}  
\delta^{s-2k}_{\mu\n}
\delta^s_{\n\n}
-
\frac{\delta^s_{\mu\mu}\delta^{s-2k-1}_{\mu \n}  
\delta^s_{\n\n}\gamma_{\mu}\gamma_{\nu}}{D+2(s-k-1)}
\right).
\ee
Up to the replacement of $\II$ by $\II^{(F)}$, the fermionic heat kernel (\ref{s9}) is the same as the 
bosonic one in eq.~(\ref{s6}). In particular, $\II^{(F)}$ carries all tensor and spinor indices of the heat kernel.\\

To evaluate the determinant of $(-\Delta^{(s+1/2)}+M^2)$ on $\mathbb{R}^D/\ZZ$, we use once more the method of 
images (\ref{t4}). As before, we need to keep track of the non-trivial index structure of 
${\cK^{AB}}_{\mu_s,\nu_s}$, which leads to
\be
\label{ss9}
\cK^{\mathbb{R}^D/\mathbb{Z}}_{\mu_s,\, \alpha_s}(t,x,x')
=
\sum_{n\,\in\,\mathbb{Z}}(-1)^n
{(J^n)_{\alpha}}^{\beta}\ldots{(J^n)_{\alpha}}^{\beta}
\,
U^n\,
\cK_{\mu_s,\,\beta_s}(t,x,\gamma^n(x')) \, ,
\ee
where the factor $(-1)^n$ comes from antiperiodic boundary conditions, $J$ is the matrix (\ref{2.7}), and $U$ 
is a $2^{\lfloor D/2\rfloor}\times2^{\lfloor D/2\rfloor}$ matrix acting on spinor indices 
and defined by
\be
\label{2.16}
{J^\alpha}_{\beta}\gamma^{\beta}
=
U \gamma^{\alpha} U^{-1}\,.
\ee
In other words, $U$ is the matrix corresponding to the transformation (\ref{2.7}) in the spinor 
representation of $\text{SO}(D)$, and it can be written as
\be
U
=
\exp\left[\frac{1}{4}\sum_{j=1}^r\theta_j[\gamma_{2j-1},\gamma_{2j}]\right].
\ee
In particular, a rotation by $2\pi$ around any given axis maps $\psi$ on $-\psi$, in accordance with the fact 
that spinors form a projective representation of $\text{SO}(D)$. Note that, using an explicit $D$-dimensional representation of the $\gamma$ matrices, one gets
\be
\label{ss9.5}
{\rm Tr}(U^n)
=
2^{\lfloor D/2\rfloor}\prod_{i=1}^r\cos(n\theta_i/2)\,.
\ee
Now, plugging (\ref{ss9}) into formula (\ref{2.5}) for the determinant of 
$-\Delta^{(s+1/2)}$, one obtains a sum of integrals which can be evaluated exactly as in the bosonic case. 
The only difference with respect to bosons comes from the spin structure, and the end result is
\be
-\log \det(-\Delta^{(s+1/2)}+M^2)
=
\sum_{n\,\in\,\mathbb{Z}^*}\frac{(-1)^n}{|n|}
\frac{\chi^{(F)}_s[n\vec\theta,\vec{\e}\,]}{\prod\limits_{j=1}^r|1-e^{in(\theta_j+i\epsilon_j)}|^2}
\times
\begin{cases} 
e^{-|n|\beta M}  &D\;{ \rm odd},\\[3pt] 
\frac{M\Delta z}{\pi}K_1(|n|\beta M)& D\;{\rm even},\end{cases}
\ee
where we have discarded a volume divergence independent of all chemical potentials (as on page 
\pageref{t5t}), and where
\be
\label{s10}
\chi^{(F)}_s[n\vec\theta,\vec{\e}\,]
=
(J^{\mu\alpha})^s\,
{\rm Tr}\left[
\II^{(F)}_{\mu_s, \alpha_s} 
\right]
\ee
is the fermionic analogue of (\ref{s5.5t}), with the same regularisation as above. This result takes exactly the same form as (\ref{ss5t}), up to 
the replacement of $\chi_s$ by $\chi_s^{(F)}$. In appendices \ref{AppB1} and \ref{AppB2}, we show that
\be
\label{s9.5}
\chi_s^{(F)}[n\vec\theta]
\stackrel{\text{\text{\ref{AppB1}}\&\ref{AppB2}}}{=}
\begin{cases}
\chi_{\lambda_s^{(F)}}[n\vec \theta\,] & \text{ for odd }D,\\[3pt]
\chi_{\lambda_s^{(F)}}[n\vec \theta,0] & \text{ for even }D,
\end{cases}
\ee
where the term on the right-hand side is the character of an 
irreducible representation of $\text{SO}(D)$ with highest weight $\lambda_s^{(F)}=(s+1/2,1/2,\ldots,1/2)$ 
(written here in the dual basis of the Cartan subalgebra of $\mathfrak{so}(D)$ described above (\ref{s6b})).\\

Having computed the required functional determinants on $\RR^D/\ZZ$, we can now write down the partition 
functions given by (\ref{2.18}) and (\ref{2.4fer}). In the massive case, the difference of Laplacians acting 
on fields with spins $(s+1/2)$ and $(s-1/2)$ produces the difference of two factors (\ref{s9.5}), with labels 
$s$ and $s-1$. It turns out that identity (\ref{t6.5b}) still holds if we replace $\lambda_s$ and 
$\lambda_{s-1}$ by their fermionic counterparts, $\lambda^{(F)}_s$ and 
$\lambda^{(F)}_{s-1}$. (The proof of this statement follows the exact same steps as in the bosonic case 
described in appendix \ref{AppA3}, up to obvious replacements that account for the change in the highest weight vector.) Accordingly, the rotating one-loop partition function of a massive field with spin 
$s+1/2$ is
\be \label{massivefermion}
Z[\beta,\vec\theta\,]
=
\exp\left[
\sum_{n=1}^\infty\frac{(-1)^n}{n}
\frac{\chi_{\lambda_s^{(F)}}^{\text{SO}(D-1)}[n\vec\theta\, \vec{\e}\,]}
{\prod\limits_{j=1}^r|1-e^{in(\theta_j+i\e_j)}|^2}
\times
\begin{cases}
e^{-n\beta M} & \text{($D$ odd)}\\
\frac{M\Delta z}{\pi}K_1(n\beta M) & \text{($D$ even)}
\end{cases}
\right].
\ee
In the massless case we must take into account one more difference of characters, namely (\ref{ss6.5t}) with 
$\lambda_s$ replaced by $\lambda^{(F)}_s$. Again, this difference can be written as a combination of 
$\text{SO}(D-2)$ characters (the proof being almost the same as in appendix \ref{AppA3}), and the partition 
function of a massless field with spin $s+1/2$ exactly takes the form (\ref{t6.5t}) or (\ref{ss6q}) (for $D$ 
even or odd, respectively) with an additional factor of $(-1)^n$ in the sum over $n$, and the replacement of 
$\lambda_s$ by $\lambda^{(F)}_s$. In particular, for $D=3$, the massless partition function can be written as
\be
\label{s9.5b}
Z
=
\prod^{\infty}_{n\,=\,s}
|1+e^{i\left(n+1/2\right)(\theta+i\epsilon)}|^2,
\ee
an expression that we will use in sect.~\ref{sec:3DSugra} and that can be recovered as 
the flat limit of the AdS result \cite{Creutzig:2011fe}. One can also verify that relation 
(\ref{s6.5q}) remains true for fermionic partition functions.

\subsection{Relation to Poincar\'e characters}\label{sec:poincare}

In this subsection we show that all one-loop partition functions displayed above can be written as 
exponentials of (sums of) Poincar\'e characters. Along the way we briefly
review the construction of induced representations of semi-direct products that will be useful also for sect.~\ref{sec:applications}. We refer e.g.\ to 
\cite{Barnich:2014kra,Oblak:2015sea} for a more 
self-contained presentation.

\subsubsection*{Representations of semi-direct products}

Let $G$ be a group, $A$ an Abelian vector group, $\sigma$ a representation of $G$ in $A$. Then 
the semi-direct product of $G$ and $A$ (with respect to $\sigma$) is the group denoted $G\ltimes_{\sigma}A$ 
(or simply $G\ltimes A$) whose elements are pairs $(f,\alpha)\in G\times A$ and whose group operation is 
$(f,\alpha)\cdot(g,\beta)=(f\cdot g,\alpha+\sigma_f\beta)$. The Poincar\'e groups are precisely of that type, 
with $G$ a Lorentz group or a spin group, and $A$ a group of translations, the action $\sigma$ then being the 
vector representation of the Lorentz group.\\

It turns out that all irreducible, unitary representations of a semi-direct product are 
induced representations analogous to 
those of the Poincar\'e group \cite{Wigner:1939,Mackey01,Mackey02}. They are classified 
by 
orbits of ``momenta'' 
belonging to the dual space of the Abelian group. Concretely, take some momentum vector $p\in A^*$ and denote by $\cO_p=\{f\cdot p\,|f\in G\}$ its 
orbit under $G$. Let 
also $\cR$ be some unitary representation of the corresponding little group $G_p$, and denote its character 
by $\chi_{\cR}$. Then, according to the Frobenius formula \cite{kirillov1976elements}, the character of the 
associated 
induced representation is
\be
\chi[(f,\alpha)]
=
\int_{\cO_p}d\mu(q)\,e^{i\langle q,\alpha\rangle}\chi_{\cR}[g_q^{-1}fg_q]\,\delta(q,f\cdot q)\,,
\label{Frob}
\ee
where $\mu$ is some (quasi-invariant) measure on $\cO_p$, $\delta$ is the associated Dirac distribution, and 
the $g_q$'s are boosts such that $g_q\cdot p=q$. One can verify that the value of $\chi[(f,\alpha)]$ is independent of 
the choice of $\mu$, and that $\chi[(f,\alpha)]$ vanishes if $f$ is not conjugate to an element of the little 
group (see e.g.\ \cite{Oblak:2015sea}). We now apply this formula to the Poincar\'e group, while in 
the next section we will use it to evaluate characters of flat $\cW_N$ algebras and of supersymmetric extensions of $\mathfrak{bms}_3$.

\subsubsection*{Poincar\'e groups and induced representations}

The connected Poincar\'e group in $D$ dimensions is a semi-direct 
product $\text{SO}(D-1,1)^{\uparrow}\ltimes\RR^D$,
where $\text{SO}(D-1,1)^{\uparrow}$ is the proper, orthochronous Lorentz group and $\RR^D$ is the group of space-time translations. The classification of (projective) irreducible, 
unitary representations of this group follows from 
the classification of orbits of momenta in terms of the value of the mass squared \cite{Wigner:1939}. In particular, massive orbits have little group $\text{SO}(D-1)$, while massless orbits have little 
group $\text{SO}(D-2)\ltimes\RR^{D-2}$. We now evaluate the characters of irreducible representations of the Poincar\'e group. To our knowledge, Poincar\'e characters have been discussed previously in \cite{joos1968,nghiem1969,Oblak:2015sea,Garbarz:2015lua}.

\subsubsection*{Massive Poincar\'e characters}

Consider a massive momentum orbit $\cO_p$ with positive energy, where $p$ is the momentum of a massive 
particle at rest, say $p=(M,0,\ldots,0)\in\RR^D$. Let $\cR$ be an 
irreducible, unitary representation of the corresponding little group $\text{SO}(D-1)$ labelled by a highest weight $\lambda$ and pick a measure $\mu$ on $\cO_p$. Since the character 
(\ref{Frob}) vanishes whenever $f$ is not conjugate to an element of the little group, we will take 
$f$ to be the rotation (\ref{2.7}). Note that we could have chosen any other rotation by the same angles 
without affecting the result, since the value of the character depends only on the conjugacy class of the 
group element at which it is evaluated. \\

When $D$ is odd, and provided all angles $\theta_1,\ldots,\theta_r$ are non-zero, the delta function 
$\delta(q,f\cdot q)$ in 
(\ref{Frob}) 
localises the integral over the orbit to a single point --- namely the momentum in the rest frame, $p$. The 
character (\ref{Frob}) then reduces to
\be
\chi[(f,\alpha)]
=
e^{iM\alpha^0}\chi_{\cR}[f]\int_{\cO_p}d\mu(q)\delta(q,f\cdot q) \, ,
\label{charBis}
\ee
where $\chi_{\cR}[f]=\chi_{\lambda}^{\text{SO}(D-1)}[\vec{\theta}\,]$ because the rotation (\ref{2.7}) belongs 
to the little group. To integrate the delta function, we use the spatial components of the momentum $q$ 
as coordinates on the orbit; in terms of these coordinates the integral reads
\be \label{intOpoincare}
\int_{\cO_p}d\mu(q)\delta(q,f\cdot q)
=
\int_{\RR^{D-1}}dq_1\ldots dq_{D-1}\delta^{(D-1)}({\bf q}-f\cdot{\bf q})
=
\prod_{j=1}^r\frac{1}{|1-e^{i\theta_j}|^2} \, ,
\ee
where we have chosen the flat Lebesgue measure on $\cO_p$ owing to the $\mu$-independence of the result.
The character (\ref{charBis}) then becomes
\be
\label{pcharMassiveOdd}
\chi[(f,\alpha)]
=
e^{iM\alpha^0}
\chi_{\lambda}^{\text{SO}(D-1)}[\vec{\theta}\,]
\prod_{j=1}^r\frac{1}{|1-e^{i\theta_j}|^2} \, .
\ee
In order to represent a particle with spin $s$, we choose the weight $\lambda$ to be $\lambda_s=(s,0,\ldots,0)$ (in the dual basis of the Cartan subalgebra of $\mathfrak{so}(D-1)$ described 
above (\ref{s6b})). With 
this choice, expression (\ref{pcharMassiveOdd}) actually appears in the exponent of (\ref{ss6.5b}): 
taking $\alpha^0=i\beta$, we can rewrite the rotating one-loop partition function for a massive field with 
spin $s$ (in odd $D$) as the exponential of a sum of Poincar\'e characters: 
\be
\label{Zexp}
Z_{M,s}[\beta,\vec\theta\,]
=
\exp\left[\,
\sum_{n\,=\,1}^{\infty}\frac{1}{n}\,\chi_{M,s}[n\vec{\theta},in\beta]
\,\right].
\ee
The series in the exponent is divergent for real $\theta_i$'s. This divergence can be regularised by 
adding suitable imaginary parts to these angles, as explained below (\ref{ss5t}).\\

In $D=3$ space-time dimensions, the massive little group is $\text{SO}(2)\cong\text{U}(1)$ and its character 
for an irreducible representation with spin $s$ is $e^{is\theta}$, so that (\ref{pcharMassiveOdd}) boils down
to
\be
\chi_{M,s}[(\text{rot}_{\theta},\alpha=i\beta)]
=
e^{-\beta M+i\theta s}\frac{1}{|1-e^{i\theta}|^2}\,,
\label{pchar3D}
\ee
where we take $\alpha$ to be a Euclidean time translation by $\beta$.\footnote{A parity-invariant version of this expression is obtained upon replacing $e^{is\th}$ by $2\cos(s\th)$.} In the next section we will show that 
the characters of flat $\cW_N$ algebras (and of supersymmetric versions of $\mathfrak{bms}_3$) are natural extensions of this 
formula.\\

When $D$ is even, the situation is more involved because the integral localises to a line instead of a 
point. For $\alpha$ being an Euclidean 
time translation by $\beta$, this leads to a non-trivial, infrared-divergent integral
\be
\int_{-\infty}^{+\infty}dk\,\delta(k-k)e^{-\beta\sqrt{M^2+k^2}}
=
\frac{M\Delta z}{\pi}K_1(\beta M)\,,
\label{pcharMassiveEven}
\ee
where we interpret $\delta(0)$ as $\Delta z/2\pi$, with $\Delta z$ the height of the rotating box along 
the space direction dual to $k$, and $K_1$ is a modified Bessel function of the second kind. Accordingly, for even $D$, the 
character of a rotation (\ref{2.7}) accompanied by a 
Euclidean time translation by $\beta$ in a massive representation of the Poincar\'e group is
\be
\label{bummer}
\chi[(f,\alpha=i\beta)]
=
\frac{M\Delta z}{\pi}K_1(\beta M)\,
\chi_{\lambda}^{\text{SO}(D-1)}[\vec{\theta}\,]
\prod_{j=1}^r\frac{1}{|1-e^{i\theta_j}|^2}\,.
\ee
For $\lambda=\lambda_s=(s,0,...,0)$, this expression 
coincides with the combinations appearing in the partition function (\ref{ss6.5b}) upon writing the latter as 
(\ref{Zexp}). The same matching works for massive fermionic fields in all space-time dimensions when replacing $\lambda_s$ by $\lambda_s^{(F)}=(s+1/2,1/2,\ldots,1/2)$.

\subsubsection*{Massless Poincar\'e characters (discrete spin)}

The little group for particles with vanishing mass is $\text{SO}(D-2)\ltimes\RR^{D-2}$. A massless 
particle is said to have {\it discrete spin} if the space of its spin 
degrees of freedom forms a finite-dimensional representation of the little group, in which all boosts spanning $\RR^{D-2}$ are represented 
trivially. In this subsection we 
will focus on such particles, relegating some comments on continuous spin particles to sect.~\ref{sec:future}. Once more, we will treat separately even and odd space-time dimensions.\\

In even space-time dimensions, any rotation in $\text{SO}(D-1)$ is conjugate to an element of
$\text{SO}(D-2)$ (in accordance with the fact that these groups have the same rank). In fact, for even 
$D$, the rotation (\ref{2.7}) belongs to the subgroup $\text{SO}(D-2)$ of 
the Lorentz group leaving invariant the momentum of a massless particle moving along the $z$ axis, so the massless characters of Poincar\'e in even 
$D$ are just the limit $M\rightarrow0$ of (\ref{pcharMassiveEven}), with the character of 
$\text{SO}(D-1)$ replaced by a 
character of $\text{SO}(D-2)$. Using also $\lim_{x\rightarrow0}xK_1(x)=1$, we 
get
\be \label{masslesschar1}
\chi[\vec\theta,\beta]
=
\frac{\Delta z}{\pi\beta}\,
\chi_{\lambda}^{\text{SO}(D-2)}[\vec{\theta}\,]
\prod_{j=1}^r\frac{1}{|1-e^{i\theta_j}|^2}\,,
\ee
which is indeed the expression appearing in the partition function (\ref{t6.5t}) 
upon writing it as (\ref{Zexp}).\\

In odd space-time dimensions, $\text{SO}(D-2)$ has lower rank than $\text{SO}(D-1)$, so the 
rotation (\ref{2.7}) is {\it not}, in general, conjugate to an element of 
the massless little group: it has one angle too much, and whenever all angles $\theta_1,\ldots,\theta_r$ 
are non-zero, the character (\ref{Frob}) vanishes. The only 
non-trivial irreducible character arises when at least one of the angles $\theta_1,\ldots,\theta_r$ vanishes, 
say $\theta_r=0$. Then the arguments explained above can be applied once more, the only subtlety 
being that the two spatial components $(k_1,k_2)$ of momentum that are not rotated produce an integral
\be
\int_{\RR^2\backslash\{0\}}dk_1dk_2\,e^{-\beta\sqrt{k_1^2+k_2^2}}\delta(k_1-k_1)\delta(k_2-k_2)
=
\frac{\Delta z\Delta z'}{2\pi\beta^2}\,,
\ee
where we have once more interpreted the delta functions evaluated at zero as 
infrared-divergent factors. The 
character of an 
irreducible massless representation of the Poincar\'e group in odd space-time dimension $D$ is thus
\be
\chi[\theta_1,\ldots,\theta_{r-1},\beta]
=
\chi^{\mathrm{SO}(D-2)}_{\lambda}[\theta_1,\ldots,\theta_{r-1}]\,
\frac{\Delta z\Delta z'}{2\pi\beta^2}
\prod_{j=1}^{r-1}\frac{1}{|1-e^{i\theta_j}|^2}\,.
\label{masslessCharIrrep}
\ee
However, comparison with (\ref{ss6q}) reveals a 
mismatch: the partition function does {\it not} take the form (\ref{Zexp}) in terms of the massless 
characters (\ref{masslessCharIrrep}); in field theory, all $r$ angles $\theta_i$ may be 
switched on simultaneously! To accommodate for this one can resort to the angle-dependent coefficients $\cA_k^r(\vec\theta)$ introduced in \eqref{s6q}. Also in this context, one can understand the origin of these coefficients through the massless limit of the 
character (\ref{pcharMassiveOdd}). Using relation (\ref{t6q}), the product of massless partition functions 
with spins ranging from zero to $s$ can be written as (\ref{Zexp}), where the characters on the right-hand 
side are massless limits of massive Poincar\'e characters. However, it is not clear to us how the quantities 
appearing in the exponent of (\ref{ss6q}) can be related directly to Poincar\'e characters {\it without} invoking a 
massless limit.

\paragraph{Remark.} The relation (\ref{Zexp}) between one-loop partition functions and characters of the 
underlying space-time isometry group is not new. 
From a  physical standpoint, it is merely the statement that a free field is a collection of 
harmonic oscillators, one for each value of momentum: the index $n$ then labels the oscillator modes, while 
the integral over momenta is the one in the Frobenius character formula (\ref{Frob}). In particular, 
standard, non-rotating one-loop partition functions are exponentials of sums of characters of (Euclidean) 
time translations. This relation has also been observed in AdS \cite{Dolan:2005wy,Gibbons:2006ij,Gopakumar:2011qs}; our partition functions are flat limits of these earlier results, up to the even-dimensional regularisation subtlety mentioned below eq.~(\ref{s5.5t}). Note that this issue already emerges at the level of characters: although most of (\ref{bummer}) is a flat limit of an $\text{SO}(D-1,2)$ character, it is not clear how to regularise the divergences that pop up when one of the angles vanishes in order to recover our regulators $\Delta z$. This problem would also appear in odd $D$ if one or more angles were set to zero. 

\section{Three-dimensional applications}\label{sec:applications}

In this section we exhibit the matching between certain combinations of higher-spin partition functions in three 
dimensions and vacuum characters of suitable asymptotic symmetry algebras. We start by reviewing the purely gravitational setting studied in \cite{Barnich:2014kra,Barnich:2015uva,Oblak:2015sea,Barnich:2015mui}, before moving on to illustrative classes of bosonic (sect.~\ref{sec:3DW}) and supersymmetric (sect.~\ref{sec:3DSugra}) higher-spin theories \cite{Afshar:2013vka,Gonzalez:2013oaa,Fuentealba:2015wza}. In the latter cases the characterisation of representations of the asymptotic symmetry algebras is subtler; nevertheless, motivated by the analogy with the gravitational setup, we propose to extend to this context several tools of the theory of induced representations recalled in sect.~\ref{sec:poincare}.

\subsection{BMS$_3$ particles and induced $\mathfrak{bms}_3$ modules}\label{BMSPart}

The asymptotic symmetries of three-dimensional gravity without cosmological constant at null infinity are given by the $\mathfrak{bms}_3$ algebra \cite{Ashtekar:1996cd,Barnich:2006av,Barnich:2010eb}, whose representations are most conveniently  analysed from the viewpoint of the underlying BMS$_3$ group. The unitary representations of this group --- or BMS$_3$ particles --- have been studied in detail in 
\cite{Barnich:2014kra,Barnich:2015uva}. Here we briefly recall some results of this analysis that we are going to extend to the higher spin setup. We also present a characterisation of induced representations at the Lie-algebraic level, that allows us to make contact with earlier proposals on the structure of unitary representations of flat $\cW$ algebras \cite{Grumiller:2014lna}.

\subsubsection*{BMS$_3$ representations}

The BMS$_3$ group is a semi-direct product $G\ltimes\mg$, where $G$ 
is the Virasoro group (consisting of superrotations) and $\mg$ is its Lie algebra, seen as an Abelian vector group\footnote{Below we will use the more 
accurate notation $G\ltimes_{\text{Ad}}\mg_{\text{Ab}}$, where $\mg_{\text{Ab}}$ is an Abelian vector 
group isomorphic to $\mg$ as a vector space, and the subscript ``$\text{Ad}$'' indicates that $G$ acts on 
$\mg_{\text{Ab}}$ according to the adjoint action.} (consisting of supertranslations):
\be
\text{BMS}_3
=
\text{Diff}(S^1)\ltimes\text{Vect}(S^1).
\label{BMS3}
\ee
Accordingly, the duals of (centrally extended) supertranslations are pairs $(p(\vf),c_2)$, where $p(\vf)$ is a supermomentum (it is a function on the circle) and $c_2$ is a central charge taking the value $c_2=3/G$ in 
Einstein gravity \cite{Barnich:2006av} (see the $\mathfrak{bms}_3$ Lie algebra (\ref{bms3})). Each induced 
representation of BMS$_3$ is associated with the orbit of such a pair $(p(\vf),c_2)$ under the action of $\text{Diff}(S^1)$, i.e.\ with a 
coadjoint orbit of the Virasoro group 
\cite{Witten:1987ty,Balog:1997zz}. The states 
of a 
BMS$_3$ particle are wavefunctions in supermomentum space, and given any (quasi-invariant) measure on the 
orbit, the corresponding
representation is unitary (see e.g.\ \cite{Oblak:2015sea}). The associated character is given by the Frobenius formula (\ref{Frob}), with the subtlety 
that the supermomentum integral is taken over an 
infinite-dimensional manifold. However, as recalled in sect.~\ref{sec:poincare}, the character vanishes whenever the element of the group which is used to determine the character in the pair $(f,\a)$ is not conjugate to an element of the little group. When $f$ is (conjugate to) a rotation by some non-zero angle $\theta$, the integral in \eqref{Frob} localises, 
so that one can compute the character explicitly \cite{Oblak:2015sea}. For a massive BMS$_3$ particle with mass 
$M$ and spin $s$, i.e.\ for a representation whose orbit contains a constant supermomentum 
$p_0=M-c_2/24$ with $M>0$, the character is given by
\be
\chi_{M,s}[(f=\text{rot}_{\theta},\alpha=i\beta)]
=
e^{-\beta M+i\theta s}\frac{1}{|1-q|^2}\cdot
e^{\beta c_2/24}\frac{1}{\prod_{n=2}^{\infty}|1-q^n|^2}\,,
\label{charBMS}
\ee
where $q=e^{i(\theta+i\epsilon)}$ with a factor $i\epsilon$ added to ensure convergence of the 
infinite product. This expression is the product of the massive Poincar\'e character 
(\ref{pchar3D}) with the vacuum BMS$_3$ character
\be
\chi_{\text{vac}}^{\text{BMS}}[(\text{rot}_{\theta},i\beta)]
=
e^{\beta c_2/24}\frac{1}{\prod_{n=2}^{\infty}|1-e^{in(\theta+i\epsilon)}|^2}\,,
\label{charSoftGrav}
\ee
which coincides with the one-loop partition 
function of gravitons given by eq.~(\ref{ss6.5q}) for $s=2$ \cite{Barnich:2015mui}.\\

The main lessons to be drawn from the previous considerations are that (i) representations of the BMS$_3$ group are labelled by orbits of supermomenta and (ii) even if the classification of such orbits requires a detailed knowledge of the finite (as opposed to infinitesimal) transformation laws of supermomenta under 
superrotations, these details are not relevant for evaluating the characters of the corresponding representations in all cases in which the integral localises. 
We focused here on representations of the BMS$_3$ group, but by differentiating them at the identity one can gain insights on the corresponding representations of the $\mathfrak{bms}_3$ algebra. By suitably generalising the notion of coadjoint orbit to the higher-spin context, in sect.~\ref{sec:3DW} we will indeed propose a (partial) classification of unitary representations of certain flat $\cW$ algebras and compute the associated characters, which are to be matched with the one-loop partition functions computed in sect.~\ref{sec:Z}. We are now going to investigate the structure of $\mathfrak{bms}_3$ representations in order to simplify the comparison between our group-inspired classification and previous proposals \cite{Grumiller:2014lna}.

\subsubsection*{Induced $\mathfrak{bms}_3$ modules}

The $\mathfrak{bms}_3$ algebra is generated by two 
infinite families of superrotation and supertranslation generators $J_m$ and $P_m$ (where $m$ is an integer) corresponding to Fourier modes of vector fields on the circle, 
together with two central charges $Z_1$ and $Z_2$. In any irreducible representation of the algebra, these charges are 
proportional to the identity operator, so 
from now on we will replace $Z_1$ and $Z_2$ 
by numbers $c_1$, $c_2$. In terms of these quantities, the commutation relations of the 
$\mathfrak{bms}_3$ algebra read \cite{Barnich:2006av,Barnich:2010eb,Barnich:2011ct}
\begin{subequations} \label{bms3}
\bea
\left[J_m,J_n\right] & = & (m-n)J_{m+n}+\frac{c_1}{12}\,m(m^2-1)\,\delta_{m+n,0}\,,\\
\left[J_m,P_n\right] & = & (m-n)P_{m+n}+\frac{c_2}{12}\,m(m^2-1)\,\delta_{m+n,0}\,,\\
\left[P_m,P_n\right] & = & 0\,. \label{PP}
\eea
\end{subequations}
Note that this algebra is a semi-direct sum
\be \label{bms3algebra}
\mathfrak{bms}_3
=
\mathfrak{vir}\inplus_{\text{Ad}}(\mathfrak{vir})_{\text{Ab}},
\ee
where $\mathfrak{vir}$ is the Virasoro algebra and $(\mathfrak{vir})_{\text{Ab}}$ denotes an Abelian Lie algebra 
isomorphic, as a vector space, to $\mathfrak{vir}$; the action of the Virasoro algebra on its Abelian counterpart is the adjoint action, as indicated by the subscript ``$\text{Ad}$''.
One way to obtain the algebra \eqref{bms3} is to take an \.In\"on\"u--Wigner contraction \cite{Inonu:1953aa} of two commuting 
copies of the Virasoro algebra, which can be physically interpreted as a flat/ultrarelativistic limit (see eq.~\eqref{weights}). One 
can indeed define
\be \label{virtobms}
P_n \equiv \frac{1}{\ell} \left(L_n + \bar{L}_{-n}\right) , \quad
J_n \equiv L_n - \bar{L}_{-n} \, , \quad
c_1 \equiv c - \bar{c} \, , \quad
c_2 \equiv \frac{c+\bar{c}}{\ell} \,,
\ee
where $L_n$ and $\bar{L}_n$ denote the generators of the two Virasoro algebras with central charges $c$ and 
$\bar{c}$ and $\ell$ is a length scale. In the limit $\ell \to \infty$ one recovers \eqref{bms3}.\\

The unitary representations of the group corresponding to the algebra \eqref{bms3} are induced, in the sense explained above. Here we wish to understand the differential of 
these representations at the identity, that is, the associated representations of $\mathfrak{bms}_3$.
For definiteness, let us consider a {\it massive BMS$_3$ particle}, whose orbit contains a constant 
supermomentum 
$p_0=M-c_2/24$ with $M>0$. 
A convenient basis of the particle's Hilbert space 
consists of plane waves, that is, wavefunctions with definite supermomentum (see e.g. section 2.3 of \cite{Barnich:2014kra}). In particular, 
there is a wavefunction whose supermomentum is the constant $p_0$, representing the 
state of the particle at rest. We will call this particular wavefunction the {\it rest-frame state} 
of the representation and denote it by $|M,s\rangle$, where $s\in\RR$ is the spin of the particle, i.e. the eigenvalue of the $J_0$ generator.
By construction, it transforms as follows under a finite supertranslation $\alpha$:
\be \label{modulesXY}
U(\alpha)|M,s\rangle=e^{i M \alpha^{0}}|M,s\rangle\,,
\quad U(\alpha)=\exp\left[i\sum_{n\,\in\,\ZZ}P_n\, \alpha^n\right].
\ee
Here $\alpha(\varphi)=\sum_{n\in\ZZ} e^{in\varphi} \alpha^n$  is a real function on the circle and $U(\alpha)$ is a unitary operator, so that $P_n^{\dagger}=P_{-n}$. By differentiating with respect to $\alpha$, one obtains 
\be
P_0|M,s\rangle=M|M,s\rangle\,,
\quad P_n|M,s\rangle=0\ \text{for}\;n\neq0\,,
\quad J_0|M,s\rangle=s|M,s\rangle\,,
\label{MM}
\ee
where the last condition comes from the definition of $|M,s\rangle$. The remaining superrotation 
generators $J_m$ (with $m\neq 0$), when acting on the rest-frame state, 
produce new states of the form
\be
J_{n_1}\cdots J_{n_k}|M,s\rangle,
\label{desc}
\ee
where $n_1,\ldots,n_k$ are arbitrary non-zero integers such that $n_1\geq\ldots\geq n_k$, and $k=0,1,2,...$ 
These additional states arise because finite superrotations act on wavefunctions as unitary operators
\be
U(\omega)
=
\exp\left[i\sum_{n\,\in\,\mathbb{Z}}J_n\,\omega^n\right]
\ee
where the complex coefficients $\omega^n=(\omega^{-n})^*$ are generalizations of the rapidity parameter of 
special relativity, and $J_n^{\dagger}=J_{-n}$. In particular, in contrast to Virasoro representations, the 
rest-frame state does not satisfy any highest-weight condition, reflecting the fact that it can be boosted in 
any direction.\\

We will call states of the form (\ref{desc}) 
{\it boosted 
states}. We also call {\it induced module} (with mass $M$ and spin $s$) the space 
$\cH_{\mathfrak{bms}}$ whose basis 
consists of the rest-frame state $|M,s\rangle$ and its boosted counterparts (\ref{desc}); it forms an irreducible representation of the 
$\mathfrak{bms}_3$ algebra (\ref{bms3}), and it is unitary by construction, since it arises from a unitary 
representation of the BMS$_3$ group. Similarly, the rest-frame state $|0\rangle$ of the vacuum induced 
module satisfies (\ref{MM}) with $M=s=0$ together with the additional condition $J_{\pm1}|0\rangle=0$, 
ensuring Lorentz-invariance. Boosted vacua are again of the form (\ref{desc}), but with all 
$n_i$'s different from $-1$, $0$ and $1$.\\

Since the algebra \eqref{bms3} emerges from the \.In\"on\"u-Wigner contraction of the conformal algebra via the redefinitions \eqref{virtobms} and the limit $\ell \to \infty$, one can also motivate the representations above by a limiting procedure. For instance, in \cite{Oblak:2015sea} it has been shown that one can recover the character \eqref{charBMS} as a flat limit of characters of non-degenerate highest-weight representations of the Virasoro algebras generated by the $L_n$ and $\bar{L}_n$ of \eqref{virtobms}. To this end, one has to write the modular parameter as $\t = \frac{1}{2\p} (\th + i\b/\ell)$ and let the highest weights $h$ and $\bar{h}$ depend on $\ell$ in such a way that the parameters
\be \label{weights}
M = \lim_{\ell \to \infty} \frac{1}{\ell} \left(h + \bar{h}\right)\,,
\qquad
s = \lim_{\ell \to \infty} \left( h - \bar{h} \right) - \frac{c_1}{24}
\ee
be finite. The conditions \eqref{MM} on the rest-frame state $|M,s\rangle$ can be seen to emerge from this limit as well, since the Virasoro highest-weight conditions translate into
\bea
\frac{1}{\ell}\,L_n |h,\bar{h}\rangle = \frac{1}{2} \left(P_n + \frac{1}{\ell}J_n \right) |h,\bar{h}\rangle =0 & \xrightarrow{\ell \to \infty} & P_n |M,s\rangle=0 \quad \text{for}\ n>0 \, , \\[5pt]
\frac{1}{\ell}\,\bar{L}_n |h,\bar{h}\rangle = \frac{1}{2} \left( P_{-n}- \frac{1}{\ell}J_{-n}  \right) |h,\bar{h}\rangle=0 & \xrightarrow{\ell \to \infty} & P_{-n} |M,s\rangle=0 \quad \text{for}\ n>0 \, ,  \, \,
\eea
i.e.\ in the second condition in \eqref{MM}, while no constraints are imposed on the $J_n$. Thus, in the 
limit one does not keep the full Virasoro highest-weight conditions, but only their leading term in a large 
$\ell$ expansion. This is analogous to what one does at the level of the algebra: after the redefinition 
\eqref{virtobms} the commutator of two $P_n$'s would be proportional to $\ell^{-2}$, but in the limit $\ell 
\to \infty$ one omits the right-hand side to get \eqref{PP}.
We stress, however, that the definition of the rest-frame state holds independently of the limit and follows 
from the theory of induced representations applied to the BMS$_3$ group. In this sense the difference between 
highest-weight and rest-frame conditions reflects the very different structure of $\mathfrak{bms}_3$ and 
Virasoro representations, generalising the difference between Poincar\'e and $\mathfrak{so}(2,2)$ 
representations. For a detailed analysis see \cite{Campoleoni:2016vsh}.

To complete the characterisation of the induced module one would like to recover the character \eqref{charBMS} by computing the trace of $e^{-\beta H+i\theta J}$ over the space defined by \eqref{MM} and \eqref{desc}. In spite of the link between \eqref{charBMS} and the flat limit of Virasoro characters recalled above, it is however not clear to us how to define a trace over induced modules that produces the desired result. This is not completely surprising since the character formula \eqref{charBMS} entails the \emph{ad hoc} regularisation obtained by adding a small imaginary part to each angle. A natural counterpart of the divergence of the infinite product in \eqref{charBMS} is the infinite multiplicity of each eigenvalue of $P_0$ and $J_0$ in the induced module, that should be regularised in some way. One should keep in mind, however, that induced modules do not capture all the features of induced representations. This is mainly due to the fact that the energy spectrum of a BMS$_3$ particle is continuous, so that, for instance, one cannot expect to be able to compute the characters of a pure supertranslation solely from the infinitesimal picture. To avoid pathologies one should stick to the characterisation of the Hilbert space of each representation in terms of wavefunctions on orbits of supermomenta rather than in terms of induced modules.\\

In spite of its limitations, the previous infinitesimal picture is useful to understand how the representations of the $\mathfrak{bms}_3$ algebra discussed in \cite{Bagchi:2009pe,Grumiller:2014lna} fit within the classification which emerges from that of induced representations of the BMS$_3$ group. As discussed above, unitary BMS$_3$ representations are labelled by orbits of pairs 
$(p(\vf),c_2)$, where $c_2$ is non-zero in Einstein gravity. Nevertheless, one may consider the 
induced representation associated with the trivial orbit of $p(\vf)=c_2=0$, whose little group is the whole 
Virasoro group generated by superrotations. In that representation, all supermomenta are set to zero, and the 
only non-trivial piece comes from the representation of the little group, which is just a standard Virasoro 
highest-weight representation obtained by starting from a highest-weight state $|s\rangle$ such that
\be
\label{s22.5}
J_0|s\rangle=s|s\rangle,\qquad J_m|s\rangle=0\ \text{ for }m>0\,.
\ee
In the Poincar\'e group, the analogue of this construction would consist in building a unitary representation 
where all translations act trivially, while Lorentz transformations are represented in a non-trivial, unitary 
way. It turns out that all unitary representations of $\mathfrak{bms}_3$ (and its higher-spin extensions) considered in \cite{Bagchi:2009pe,Grumiller:2014lna} were of this type. The authors attempted to build 
representations by enforcing the conditions (\ref{s22.5}) while replacing $|s\rangle$ by a state 
$|M,s\rangle$, with $M$ the energy of the state. Upon switching on the central charge $c_2$, they concluded 
that unitary representations arise {\it only} if $M=c_2=0$. But as we can see from our earlier 
considerations, this had to be so: the highest-weight conditions (\ref{s22.5}) rely on the assumption that 
superrotations are represented as in a usual CFT, which occurs only for $M=c_2=0$. By contrast, for non-zero 
$c_2$, the suitable conditions are not (\ref{s22.5}), but the rest-frame conditions (\ref{MM}).

\subsection{Characters of flat $\cW_N$ algebras}\label{sec:3DW}

We now move to the asymptotic symmetry algebras that arise at null infinity for higher-spin theories in flat space. We propose a way to characterise their unitary representations and compute the associated characters, showing in 
particular that vacuum characters match certain combinations of the one-loop partition functions displayed in 
sect.~\ref{subsec2.2}. The coadjoint representation of standard $\cW_N$ algebras \cite{Balog:1990mu,Khesin:1991ra,Bajnok:2000nb} plays 
a key role in our analysis, so we start by first reviewing results from the AdS context.

\subsubsection*{Higher spins in AdS$_3$ and the $\cW_3$ algebra}

Asymptotic symmetries of higher-spin theories in three dimensions were first studied in 
AdS$_3$ \cite{Henneaux:2010xg,Campoleoni:2010zq,Gaberdiel:2011wb,Campoleoni:2011hg}, where they typically span 
the direct sum of two non-linear $\cW$ algebras. Here we focus on models including fields with spin ranging 
from 2 to $N$.\footnote{In AdS$_3$ this setup can be described by an 
$\mathfrak{sl}(N,\RR)\oplus\mathfrak{sl}(N,\RR)$ Chern-Simons action with a principally embedded 
$\mathfrak{sl}(2,\RR)\oplus\mathfrak{sl}(2,\RR)$ gravitational subalgebra.} When 
$N=3$, the asymptotic symmetries are generated by gauge transformations specified by 
four arbitrary, $2\pi$-periodic functions $(X(x^+),\xi(x^+))$ and $(\bar X(x^-),\bar\xi(x^-))$ of the 
light-cone coordinates $x^{\pm}$ on the boundary of AdS$_3$. In particular, the functions 
$X(x^+)$ and $\bar X(x^-)$ generate conformal transformations \cite{Brown:1986nw,Campoleoni:2014tfa}. Since the results are 
left-right symmetric, we focus on 
the left-moving sector. The surface charge associated with a transformation $(X,\xi)$ then takes 
the form \cite{Campoleoni:2010zq}
\be
Q_{(X,\xi)}[p,\rho]
=
\frac{1}{2\pi}\int_0^{2\pi}d\varphi\left[X(\vf)p(\vf)+\xi(\vf)\rho(\vf)\right],
\label{chargeW}
\ee
where $\varphi=(x^+-x^-)/2$, while $p(\vf)$ and $\rho(\vf)$ are two arbitrary, $2\pi$-periodic functions 
specifying a solution of the field equations at fixed time. In fact, if we think of the pair $(X,\xi)$ as being an element 
of the $\cW_3$ algebra, the charge (\ref{chargeW}) is the pairing between $\cW_3$ and 
its dual space. Accordingly, $(p,\rho)$ may be seen as a coadjoint vector of the $\cW_3$ algebra. Its 
infinitesimal transformation law is given by \cite{Campoleoni:2010zq}
\begin{subequations}\label{deltaP}
\bea
\delta_{(X,\xi)}p
& = &
Xp'+2\,X'p-\frac{c}{12}\,X'''+2\,\xi\rho'+3\,\xi'\rho\,,\\[5pt]
\delta_{(X,\xi)}\rho
& = &
X\rho'+3\,X'\rho+\frac{\sigma}{3}\Big[-2\,\xi p'''-9\,\xi'p''-15\,\xi''p'-10\,\xi'''p+\nn\\
\label{deltaPi}
&   & +\,\frac{c}{12}\,\xi^{(5)}+\frac{192}{c}\left(\xi\,pp'+\xi'p^2\right)\Big]\,,
\eea
\end{subequations}
where $\sigma$ is an irrelevant normalisation factor, prime 
denotes differentiation with respect to $x^+$, 
and $c=3\ell/2G$ is the Brown-Henneaux central charge \cite{Brown:1986nw} (with $\ell$ the AdS radius). The infinitesimal transformations generated by $X$ imply that $p$ is a quasi-primary field with weight 2 with respect to conformal transformations, 
while $\rho$ is a primary field with weight 3. Together with the surface charges (\ref{chargeW}), these 
transformation laws yield the Poisson 
bracket
\be
\label{ss25}
\left\{Q_{(X,\xi)}[p,\pi],Q_{(Y,\zeta)}[p,\pi]\right\}
=
-\,\delta_{(X,\xi)}Q_{(Y,\zeta)}[p,\pi]\,,
\ee
which coincides with the non-linear bracket of a $\cW_3$ algebra with central charge 
$c$. Similar considerations apply to models including fields with spin ranging from 2 to $N$ \cite{Campoleoni:2010zq,Campoleoni:2011hg}. The 
resulting asymptotic symmetry algebra is the direct sum of two copies of $\cW_N$.

\subsubsection*{Flat $\cW_3$ algebra}

The asymptotic symmetries of higher-spin theories at null infinity in three-dimensional flat space were 
discussed in 
\cite{Afshar:2013vka,Gonzalez:2013oaa,Grumiller:2014lna}. For the model describing the gravitational coupling of a field of spin 3,\footnote{Its action can be still written in a Chern-Simons form, with the gauge algebra $\mathfrak{sl}(3,\RR) \inplus (\mathfrak{sl}(3,\RR))_{\text{Ab}}$. The latter can be obtained as the \.In\"on\"u-Wigner contraction of the $\mathfrak{sl}(3,\RR)\oplus\mathfrak{sl}(3,\RR)$ gauge algebra of the corresponding model in AdS$_3$ \cite{Afshar:2013vka}.} it was 
found that symmetry transformations are labelled by four arbitrary, $2\pi$-periodic functions 
$X(\varphi)$, $\xi(\varphi)$, $\alpha(\varphi)$ and $a(\varphi)$ on the celestial circle at null infinity. Of these, $X(\varphi)$ and $\alpha(\vf)$ generate standard BMS$_3$ superrotations and supertranslations 
(respectively), while $\xi$ and $a$ are their higher-spin extensions. The corresponding surface charges read
\be
\label{s25}
Q_{(X,\xi,\alpha,a)}[j,\kappa,p,\rho]
=
\frac{1}{2\pi}
\int_0^{2\pi}d\varphi
\left[
X(\varphi)j(\varphi)+\xi(\varphi)\kappa(\varphi)+\alpha(\varphi)p(\varphi)+a(\varphi)\rho(\varphi)
\right],
\ee
where the $2\pi$-periodic functions $j$, $\kappa$, $p$ and $\rho$ determine a solution of the equations of 
motion. $p(\varphi)$ is the standard Bondi mass aspect (supermomentum), while $j(\varphi)$ is the 
angular momentum aspect (angular supermomentum); the functions $\rho$ and $\kappa$ generalise these 
quantities 
for a spin-3 field. As in the AdS case, the quadruple $(j,\kappa,p,\rho)$ may be seen as an element of the 
dual space of the asymptotic symmetry algebra. In particular, the higher-spin supermomentum $(p,\rho)$ 
transforms under 
higher-spin superrotations $(X,\xi)$ as a coadjoint vector of the $\cW_3$ algebra, that is, according to 
(\ref{deltaP}), albeit with a central charge $c_2=3/G$ instead of $c=3\ell/2G$.\\

Inspection of the Poisson brackets satisfied by the surface charges (\ref{s25}), as displayed for instance in 
\cite{Gonzalez:2013oaa,Afshar:2013vka}, reveals that, in analogy with \eqref{bms3algebra}, the asymptotic symmetry algebra is a semi-direct sum
\be
\text{``flat $\cW_3$ algebra''}
\equiv
\cF\cW_3
=
\cW_3\inplus_{\text{Ad}}(\cW_3)_{\text{Ab}}\,,
\label{Ww}
\ee
where $\cW_3$ is the standard $\cW_3$ algebra and $(\cW_3)_{\text{Ab}}$ denotes an Abelian Lie algebra 
isomorphic, as a vector space, to $\cW_3$. This algebra is centrally extended, as the bracket 
between generators of $\cW_3$ and those of $(\cW_3)_{\text{Ab}}$ includes a central charge $c_2$.

\subsubsection*{Induced representations, unitarity and characters}

Since the flat $\cW_3$ algebra (\ref{Ww}) has the form $\mg\inplus\mg_{\text{Ab}}$, with $\mg$ the standard 
$\cW_3$ algebra, its unitary representations should be induced representations labelled by orbits of 
supermomenta under the coadjoint action of elements of a groupoid whose differentiation gives $\cW_3$. 
However, the non-linearities that appear in $\cW$ algebras make this step subtle. In the cases where 
the definition of the group is under control, as for $\text{BMS}_3$, acting with group elements is required 
to specify the finite transformation of the supermomenta. This characterises the full orbit on which to 
define the wavefunctions that give a basis of the Hilbert space of each representation. Fortunately, one can 
bypass the need to control the group as follows. Generic $\cW$ algebras define a Poisson manifold through 
\eqref{ss25} and one can classify the submanifolds on which the Poisson structure is invertible, called 
\emph{symplectic leaves} \cite{Khesin:1991ra}. In the case of the Virasoro algebra (which corresponds to the 
$\cW_N$ algebra with $N=2$) this concept coincides with that of a coadjoint orbit of the Virasoro group. We 
thus propose to build unitary representations of flat $\cW_N$ algebras as Hilbert spaces of wavefunctions 
defined on their symplectic leaves, on which we assume that one can define a suitable (quasi-invariant) 
measure. (See e.g.~\cite{ShavgulidzeBis} for the construction of such a measure in the case of the Virasoro 
group.) One can make the analogy between symplectic leaves and coadjoint orbits even stronger: symplectic 
leaves of $\cW_N$ algebras can be obtained as intersections of the coadjoint orbits of $\mathfrak{sl}(N)$-Kac 
Moody algebras with the constraints that implement the Hamiltonian reduction to $\cW_N$ algebras 
\cite{Bajnok:2000nb}.\\

A complete classification of the symplectic leaves of the standard $\cW_3$ algebra has been worked out in \cite{Khesin:1991ra,Bajnok:2000nb} and, according to our proposal, this provides the basis for a complete classification of irreducible, unitary representations of the flat $\cW_3$ algebra. Here, following \cite{Witten:1987ty}, we restrict instead our analysis to orbits of constant supermomenta, which can be classified from the infinitesimal transformation laws \eqref{deltaP} given by the algebra. 
To describe the orbits of constant supermomenta let us pick a pair $(p,\rho)$ where $p(\varphi)=p_0$ and $\rho(\varphi)=\rho_0$ are constants, and act on it with an infinitesimal higher-spin superrotation $(X,\xi)$. Then, all terms involving derivatives of $p$ or 
$\rho$ in the transformation law (\ref{deltaP}) vanish, and we find
\begin{subequations}\label{deltaP0}
\bea
\delta_{(X,\xi)}p_0
& = &
2\,X'p_0-\frac{c_2}{12}\,X'''+3\,\xi'\rho_0\,,\\[5pt]
\delta_{(X,\xi)}\rho_0
& = &
3\,X'\rho_0+\frac{\sigma}{3}\left[\,-10\,\xi'''p_0+\frac{c_2}{12}\,\xi^{(5)}+\frac{192}{c_2}\,\xi'p_0^2\,\right]\label{deltaPi0}.
\eea
\end{subequations}
The little group for $(p_0,\rho_0)$ consists of higher-spin superrotations leaving it invariant. 
The little algebra is therefore spanned by pairs $(X,\xi)$ such that the right-hand sides of 
eqs.~(\ref{deltaP0}) vanish:
\begin{subequations}\label{lilp}
\bea
2\,X'p_0-\frac{c_2}{12}X'''+3\,\xi'\rho_0 & = & 0\,,\\
\label{lilpi}
3\,X'\rho_0+\frac{\sigma}{3}\left[\,-10\,\xi'''p_0+\frac{c_2}{12}\,\xi^{(5)}+\frac{192}{c_2}\,\xi'p_0^2\,\right] & 
= & 0\,.
\eea
\end{subequations}
The solutions of these equations depend on the values of $p_0$ and $\rho_0$. Here we will 
take $\rho_0=0$ for simplicity, i.e.~we only consider cases where all higher-spin charges are switched off. 
\label{Off} Then, given $p_0$, eqs.~(\ref{lilp}) become two decoupled differential 
equations for the functions $X(\vf)$ and $\xi(\vf)$, leading to three different cases:
\begin{itemize}
\item For generic values of $p_0$, the only pairs $(X,\xi)$ leaving $(p_0,0)$ invariant are constants, and 
generate a little group $\text{U}(1)\times\RR$. 
\item For $p_0=-n^2c_2/96$ where $n$ is a positive odd integer, the pairs $(X,\xi)$ leaving $(p_0,0)$ 
invariant are 
of the form
\be
X(\varphi)=A,\quad
\xi(\varphi)=B+C\cos(n\vf)+D\sin(n\vf),
\ee
where $A$, $B$, $C$ and $D$ are real numbers. The corresponding little group is the $n$-fold cover of 
$\text{GL}(2,\RR)$.
\item For $p_0=-n^2c_2/24=-(2n)^2c_2/96$ where $n$ is a positive integer, the Lie algebra of the 
little group is spanned by
\be
\begin{split}
X(\varphi)= &
\,A+B\cos(n\vf)+C\sin(n\vf),\\
\xi(\varphi)= &
\,D+E\cos(n\vf)+F\sin(n\vf)+G\cos(2n\vf)+H\sin(2n\vf),
\end{split}
\label{exlil}
\ee
where $A,B,...,H$ are real coefficients. The little group is thus an $n$-fold cover of 
$\text{SL}(3,\mathbb{R})$. In particular, $p_0=-c_2/24$ realises the absolute minimum of energy among all 
supermomenta belonging to orbits with energy bounded from below. It is thus the supermomentum of the vacuum 
state, and indeed, upon using $c_2=3/G$, the field configuration that corresponds to it is the metric of 
Minkowski space (with the spin-3 field set to zero on account of $\rho_0=0$).\\
\end{itemize}

The previous information on little groups is actually sufficient to evaluate certain characters along the lines of \cite{Oblak:2015sea}. 
For instance, consider an induced module based on the orbit of a generic pair $(p_0,0)$, and call $(s,\sigma)$ the spins of the 
representation $\cR$ of the little group 
$\text{U}(1)\times\RR$. Then take a superrotation which is an element of the $\text{U}(1)$ subalgebra (i.e.\ a rotation 
$f(\varphi)=\vf+\theta$), and whose higher-spin supertranslation is an arbitrary 
combination 
$(\alpha(\varphi),a(\varphi))$. The only point on the orbit that is left invariant by the rotation 
is $(p_0,0)$, and the whole integral over the orbit in (\ref{Frob}) localises to that 
point. Therefore, in analogy with the BMS$_3$ example, the detailed knowledge of the orbit is irrelevant to compute the character. In particular, the only components of 
$\alpha(\varphi)$ and $a(\varphi)$ that survive the integration are their zero-modes $\alpha^0$ and $a^0$, 
and 
the character takes the form 
\be
\chi[(\text{rot}_{\theta},\alpha,a)]
=
e^{is\theta}e^{ip_0\alpha^0}\int_{\cO_{p_0}}\!
\cD\mu(q)\,\delta(q,\text{rot}_{\theta}\cdot q)\,.
\ee
In writing this we assumed the existence of a (quasi-invariant) measure $\mu$ on the orbit, whose precise expression is unimportant since different measures give representations that are unitarily equivalent \cite{Oblak:2015sea}. We have manifested that the little group character reduces to $e^{is\theta}$, so that this expression is an infinite-dimensional counterpart of \eqref{charBis}. Our remaining task is to integrate the delta 
function. To do so, we use local coordinates on the orbit, which we choose to be the Fourier modes of higher-spin 
supermomenta in analogy with \eqref{intOpoincare}. Because $p_0$ is generic, the non-redundant coordinates on the 
orbit are the non-zero modes. The integral is thus
\be
\int_{\cO_{p_0}}\!\!
\cD\mu(q)\delta(q,\text{rot}_{\theta}\cdot q)
=
\prod_{n\,\in\,\mathbb{Z}^*}\!\left(\int dq_n\delta(q_n-e^{in\theta}q_n)\right)\!
\prod_{m\,\in\,\mathbb{Z}^*}\!\left(\int d\rho_m\delta(\rho_m-e^{im\theta}\rho_m)\right),
\label{intFourier}
\ee
where we call $q_n$ the Fourier modes of the standard (spin 2) supermomentum, while $\rho_m$ are the modes of 
its higher-spin counterpart. Performing the integrals over Fourier modes and adding small 
imaginary parts $i\epsilon$ to $\theta$ to ensure convergence of the character, one obtains
\be
\chi[(\text{rot}_{\theta},\alpha,a)]
=
e^{is\theta}e^{ip_0\alpha^0}
\left(\prod_{n=1}^{\infty}\frac{1}{|1-e^{in(\theta+i\epsilon)}|^2}
\right)^2.
\ee
This is a natural spin-3 extension of the spin-2 (BMS$_3$) massive character 
(\ref{charBMS}).\\

A similar computation can be performed for orbits of other higher-spin supermomenta $(p_0,0)$. The only 
subtlety is that, for the values of $p_0$ for which the little 
group is larger than $\text{U}(1)\times\RR$, the orbit has higher codimension in $\cW_3^*$ than the 
generic orbit we just discussed. Accordingly, there are fewer coordinates on the orbit and the 
products of integrals (\ref{intFourier}) are truncated. For instance, when $p_0=-n^2c/24$ with $n$ a positive integer, the little group is generated by pairs 
$(X,\xi)$ of the form (\ref{exlil}), so that the Fourier modes providing non-redundant local coordinates on 
the orbit (in a neighbourhood of $(p_0,0)$) are the modes $q_m$ with $m\notin\{-n,0,n\}$ and the higher-spin 
modes $\rho_m$ with $m\notin\{-2n,-n,0,n,2n\}$. Assuming that the representation $\cR$ of the little group is trivial, this produces a character
\be
\chi[(\text{rot}_{\theta},\alpha,a)]
=
e^{-in^2c_2\alpha^0/24}
\Bigg(
\prod_{\substack{m=1,\\ m\neq n}}^{\infty}
\frac{1}{|1-e^{im(\theta+i\epsilon)}|^2}
\Bigg)
\cdot
\Bigg(
\prod_{\substack{m=1,\\ m\neq n,\\ m\neq2n}}^{\infty}
\frac{1}{|1-e^{im(\theta+i\epsilon)}|^2}
\Bigg)
\,.
\ee
The choice $n=1$ specifies the vacuum representation of the flat 
$\cF\cW_3$ algebra; taking $\alpha$ to be a Euclidean time translation by $i\beta$, we get
\be \label{vac3}
\chi_{\text{vac}}[(\text{rot}_{\theta},\alpha=i\beta,a=0)]
=
e^{\beta c_2/24}
\Bigg(
\prod_{n=2}^{\infty}
\frac{1}{|1-e^{in(\theta+i\epsilon)}|^2}
\Bigg)
\cdot
\Bigg(
\prod_{n=3}^{\infty}
\frac{1}{|1-e^{in(\theta+i\epsilon)}|^2}
\Bigg).
\ee
Comparing with eq.~(\ref{ss6.5q}), we recognise the product of the rotating one-loop partition functions 
of massless fields with spins two and three in three-dimensional flat space, including the classical piece 
$S^{(0)}=-\beta c_2/24$.\footnote{The value of the classical contribution depends on the normalisation of the Bondi mass. For example, if the normalisation is chosen in such a way  that the mass of Minkowski space-time vanishes, then the corresponding classical action would vanish, as would the exponential prefactor of the character. } 
This is one of our key results, that provides a first non-trivial check of our proposal to construct unitary representations of flat $\cW_N$ algebras.\\

All the induced representations described above are unitary by construction, provided one can 
define (quasi-invariant) measures on the corresponding orbits. In analogy with representations of the 
$\mathfrak{bms}_3$ algebra, they can also be described in terms of induced modules that generalise those 
discussed on page \pageref{modulesXY}. Accordingly, one can again define a rest-frame state as one that is 
annihilated by all non-zero Fourier modes of the supertranslation generators $p$ and $\r$ introduced in 
\eqref{s25}. Boosted states are obtained by acting with \emph{all} Fourier modes of 
the superrotation generators $j$ and $\k$. Our representations thus evade the no-go theorems of 
\cite{Grumiller:2014lna} that stated the absence of unitary representations of the algebra \eqref{Ww} under 
certain conditions. The reason is that the representations considered in \cite{Grumiller:2014lna} are higher 
spin generalisations of those described in \eqref{s22.5}, and as such required $c_2=0$. Since some of the 
non-linear terms in (\ref{deltaPi0}) depend on inverse powers of $c_2$ one had to first properly rescale some 
of the generators before taking $c_2\rightarrow0$, which in turn rendered all higher-spin excitations to be 
null states, thus resulting in unitary representations of 
$\mathcal{FW}_3$ without higher-spin states. This argument, however, does not apply to the induced 
representations considered in this paper as these representations are unitary and allow for $c_2\neq0$ without 
rendering all the higher-spin states unphysical.

\subsubsection*{Flat $\cW_N$ algebras}

The considerations of the previous pages can be generalised to higher-spin theories in flat 
space with spins ranging from 2 to $N$. In AdS$_3$ the asymptotic symmetries of models with this field content 
are given by two copies of a $\cW_N$ algebra and it is natural to anticipate that the corresponding theory in 
flat space $N$ will have an asymptotic symmetry algebra
\be
\text{``flat $\cW_N$ algebra''}
\equiv
\cF\cW_N
=
\cW_N\inplus_{\text{Ad}}(\cW_N)_{\text{Ab}}\,,
\label{s29.5}
\ee
in analogy with \eqref{bms3algebra} and \eqref{Ww}.
The surface charges generating these symmetries should coincide with the pairing of the Lie algebra of 
(\ref{s29.5}) with its dual space, and they should satisfy a centrally extended algebra. Since the presence 
of higher-spin fields does not affect the value of the central charge in three-dimensional AdS gravity \cite{Henneaux:2010xg,Campoleoni:2010zq}, we expect the central charge in this case to be the usual $c_2=3/G$ 
appearing in mixed brackets \cite{Barnich:2006av}. This structure was indeed observed for $N=4$ in 
\cite{Grumiller:2014lna}. We will now argue that this proposal must hold for any $N$ by showing that the 
vacuum character of (\ref{s29.5}), computed along the lines followed above for $\cF\cW_3$, reproduces the 
product of one-loop partition functions of fields of spin $2,3,\ldots,N$.\\

According to our proposal for the characterisation of the representations of semi-direct sums of the type (\ref{s29.5}), unitary representation of flat $\cW_N$ algebras are classified by their symplectic leaves, that is, by orbits of higher-spin supermomenta $(p_1,\ldots,p_{N-1})$. (Here $p_1(\vf)$ is the supermomentum that we used to write as $p(\vf)$, while 
$p_2(\vf)$ is what we called $\rho(\vf)$ for $N=3$.) The infinitesimal transformations that generalise \eqref{deltaP} and that define these orbits locally can 
be found e.g.\ in \cite{Campoleoni:2011hg}. Here we focus on the vacuum orbit where we set all higher-spin 
charges to zero and take only $p_1=-c_2/24$ to be non-vanishing. This particular supermomentum is left fixed 
by higher-spin asymptotic symmetries of the form
\be\label{eq:LastEqFinal}
X_i(\vf)
=
A_i+
\sum_{j=1}^i\left(B_{ij}\cos(j\vf)+C_{ij}\sin(j\vf)\right),
\quad i=1,\ldots,N-1,
\ee
where the coefficients $A_i$, $B_{ij}$, $C_{ij}$ are real. In principle, one can obtain such symmetry 
generators by looking for the stabiliser of the vacuum as in \eqref{lilp}, using for instance the explicit 
formulas of \cite{Campoleoni:2011hg}. However, a much simpler way to derive the same result is to look for the 
higher-spin isometries of the vacuum in the Chern-Simons formulation of the dynamics, in which models with 
fields of spin ranging from $2$ to $N$  are described in flat space by a Chern-Simons action with gauge 
algebra $\mathfrak{sl}(N,\mathbb{R})\inplus (\mathfrak{sl}(N,\mathbb{R}))_{\text{Ab}}$ (see e.g.\ 
\cite{Campoleoni:2011tn,Gonzalez:2013oaa,Grumiller:2014lna}). In retarded Bondi coordinates $(r,u,\vf)$, the 
vacuum field configuration takes the form
\be
A_{\mu}(x)
=
b(r)^{-1}g(u,\vf)^{-1}
\partial_{\mu}
\left[g(u,\vf)b(r)\right],
\qquad
b(r)=
\exp\left[\frac{r}{2}\,P_{-1}\right],
\ee
where $g(u,\vf)$ is the $\text{SL}(N,\RR)\ltimes\mathfrak{sl}(N,\RR)$ valued field given by
\be
\label{ggg}
g(u,\vf)=
\exp\left[
\left(P_1+\frac{1}{4}\,P_{-1}\right)u+
\left(J_1+\frac{1}{4}\,J_{-1}\right)\varphi\,
\right]
\ee
in terms of generators of the Poincar\'e algebra satisfying the commutation relations (\ref{bms3}) with 
$m,n=-1,0,1$ (and of course without central extensions). The isometries of this field configuration are 
generated by gauge parameters of the form $(g\cdot b)^{-1}T_a(g\cdot b)$, where $T_a$ is any of the basis 
elements of the gauge algebra. Upon expanding $g^{-1}T_ag$ as a position-dependent linear combination of gauge 
algebra generators, the function multiplying the lowest weight generator coincides with the corresponding 
asymptotic symmetry parameter (see e.g.\ \cite{Campoleoni:2010zq} for more details). The latter can be 
obtained as follows.\\

For convenience, let us diagonalise the Lorentz piece of the group element (\ref{ggg}) as
\be
\exp\left[
\left(J_1+\frac{1}{4}J_{-1}\right)\varphi
\right]
=Be^{iJ_0\varphi}B^{-1}
\ee
where $B$ is some $\text{SL}(2,\RR)$ matrix. Then the gauge parameters generating the little group of the vacuum configuration can be written as
\begin{subequations}
\bea
& & \exp\left[-\left(J_1+\frac{1}{4}J_{-1}\right)\varphi
\right]
\sum_{m=-\ell}^{\ell}\alpha^mW_m^{(\ell)}
\exp\left[
\left(J_1+\frac{1}{4}J_{-1}\right)\varphi
\right]\label{fgfg}\\
& = &
\label{twiddle}
B\,e^{-iJ_0\vf}\,\sum_{m=-\ell}^{\ell}
\alpha^m\,B^{-1}\,W_m^{(\ell)}\,B\,e^{iJ_0\vf}\,B^{-1},
\eea
\end{subequations}
where the $\alpha^m$'s are certain real coefficients, while the $W_m^{(\ell)}$ (with $2 \leq \ell \leq N$ and 
$- \ell \leq m \leq \ell$) are the generators of the $\mathfrak{sl}(N,\mathbb{R})$ algebra (including the 
$J_m \equiv W^{(2)}_m$). Note that the matrix $B$ preserves the conformal weight since it is an exponential 
of $\mathfrak{sl}(2,\RR)$ generators, so that 
\be
\sum_{m=-\ell}^{\ell}\alpha^m\,B\,W_m^{(\ell)}\,B^{-1}
=
\sum_{m=-\ell}^{\ell}\tilde\alpha^mW_m^{(\ell)}
\ee
for some coefficients $\tilde\alpha^j$ obtained by acting on the $\alpha^m$'s with an invertible linear map. Because each generator $W_m^{(\ell)}$ has weight $m$ under $J_0$, expression (\ref{twiddle}) can be rewritten as 
\be
\sum_{m={-\ell}}^{\ell}
e^{im\vf}
\tilde\alpha^m\,
B\,W_m^{(\ell)}\,B^{-1}
=
\sum_{m,n={-\ell}}^{\ell}\beta^{mn}W_n^{(\ell)}
e^{ij\vf}
=
\sum_{m=-\ell}^{\ell}e^{im\vf}\beta^{m\ell}W_{\ell}^{(\ell)}
+
\cdots
\ee
for some coefficients $\beta^{mn}$. In the last step we omitted all terms proportional to $W_m^{(\ell)}$'s with $m<\ell$; the important piece is the term that multiplies the highest-weight generator $W_{\ell}^{(\ell)}$: it is the function on the circle that generates the asymptotic symmetry corresponding to the generator $\sum_{m=-\ell}^{\ell}\alpha^mW_m^{(\ell)}$ that we started with in (\ref{fgfg}). Since the $\beta^{m\ell}$'s are related to the $\alpha^m$'s by an invertible linear map, and since there are $2\ell+1$ linearly independent generators of this type, the isometries of the vacuum exactly span the set of functions of the form (\ref{eq:LastEqFinal}). This is what we wanted to prove; there are $N^2-1$ linearly independent asymptotic symmetry generators of this form, and they span the Lie algebra of $\text{SL}(N,\RR)$.\\

The character associated with the vacuum 
representation of (\ref{s29.5}) can then be worked out exactly as in the cases $N=2$ and $N=3$ discussed 
above: using the Fourier modes of the $N-1$ components of supermomentum as coordinates on the orbit, we need 
to mod out the redundant modes. For the vacuum orbit, these are the modes ranging from $-\ell$ to $\ell$ for the $\ell^{\text{th}}$ component. The integral over the localising delta function in the Frobenius formula 
(\ref{Frob}) then produces a character
\be
\chi[(\text{rot}_{\theta},a_1=i\beta,a_2=\ldots=a_{N-1}=0)]
=
e^{\beta c_2/24}
\prod_{s=2}^N
\left(\prod_{n=s}^{\infty}\frac{1}{|1-e^{in(\theta+i\epsilon)|^2}}\right).
\ee
Comparing with (\ref{ss6.5q}), we recognise the product of one-loop partition functions of massless higher-spin fields 
with spins ranging from 2 to $N$, including a classical contribution. This result confirms, on the one hand, our conjecture (\ref{s29.5}) for the asymptotic symmetry algebras of a generic higher-spin theory in three-dimensional flat 
space, and on the other hand it provides another consistency check of our proposal for the characterisations of the unitary representations of flat $\cW_N$ algebras.

\subsection{Supersymmetry and super BMS$_3$ characters}\label{sec:3DSugra}

The supersymmetric BMS$_3$ groups describe the symmetries of three-dimensional, asymptotically flat 
supergravity \cite{Barnich:2014cwa,Barnich:2015sca,Fuentealba:2015jma,Fuentealba:2015wza,Mandal:2010gx}. Here 
we exhibit the classification of unitary 
representations of their $\cN=1$ version and show that the corresponding vacuum character 
coincides (in the Neveu-Schwarz sector) with the one-loop partition function of $\cN=1$ supergravity. We then extend the matching between vacuum characters and one-loop partition functions to hypergravity theories, describing the gravitational coupling of a massless field of spin $s+1/2$. We 
start by reviewing briefly unitary representations of supersymmetric 
semi-direct products, referring to \cite{Carmeli:2005rg,CarmeliThesis} for details.

\subsubsection*{Supersymmetric induced representations}\index{notasubsec}

A super Lie group is a pair
$(\Gamma_0,\gamma)$ 
where $\Gamma_0$ is a Lie group in the 
standard sense, while $\gamma$ is a super Lie algebra whose even part coincides with the Lie algebra of 
$\Gamma_0$, and whose odd part is a $\Gamma_0$-module such that the differential of the $\Gamma_0$ action be 
the bracket 
between even and odd elements of $\gamma$ \cite{deligne1999quantum}. Then a {\it super 
semi-direct product} is a super Lie group of the form \cite{Carmeli:2005rg,CarmeliThesis}
\be
\left(
G\ltimes_{\sigma}A
,
\mathfrak{g}\inplus(A+\cA)
\right),
\label{supergroup}
\ee
where $G\ltimes A$ is a standard (bosonic) semi-direct product group with Lie algebra $\mathfrak{g}\inplus 
A$, and $\mathfrak{g}\inplus(A+\cA)$ is a super Lie algebra whose odd subalgebra $\cA$ is a $G$-module such that
the bracket between elements of $\mathfrak{g}$ and elements of $\cA$ be the differential of the action of $G$ 
on $\cA$, and such that $[A,\cA]=0$ and $\{\cA,\cA\}\subseteq A$. By virtue of this definition, the action 
$\phi$ of $G$ on $\cA$ is compatible with the super Lie bracket:
\be
\left\{\phi_gS,\phi_gT\right\}
=
\sigma_g\left\{S,T\right\}
\quad\forall\,S,T\in\cA,
\label{compa}
\ee
where $\sigma$ is the action of $G$ on $A$.\\

It was shown in \cite{Carmeli:2005rg,CarmeliThesis} that all irreducible, unitary representations 
of a super 
semi-direct product are induced in essentially the same sense as for standard, bosonic groups. In particular, 
they are classified by the orbits and little groups of $G\ltimes_{\sigma}A$. However, there are two important differences with 
respect to the purely bosonic case:
\begin{enumerate}
\item Unitarity rules out all orbits on which energy can be negative, so that the
momentum orbits giving rise to unitary representations of the supergroup form a subset of the full menu of 
orbits available in the purely bosonic case. More precisely, 
given a momentum $p\in A^*$, it must be such that
\be
\langle p,\{S,S\}\rangle\geq0\quad\forall S\in\cA.
\label{admiss}
\ee
If this condition is not satisfied, the representations of (\ref{supergroup}) 
associated 
with the orbit $\cO_p$ are not unitary. The momenta satisfying condition (\ref{admiss}) are said to be 
{\it admissible}. Note that admissibility is a $G$-invariant statement: if $f\in G$ and if $p$ is admissible, 
then so is $f\cdot p$, by virtue of (\ref{compa}). For example, for Poincar\'e, the only admissible momenta 
are those of massive or massless particles with positive energy (and the trivial momentum $p=0$).
\item Given an admissible momentum $p$, the odd piece $\cA$ of the supersymmetric translation algebra 
produces a (generally degenerate) Clifford algebra
\be
\cC_p
=
T(\cA)/\left\{S^2-\langle p,\{S,S\}\rangle\;|\;S\in\cA\right\},
\label{cliff}
\ee
where $T(\cA)$ is the tensor algebra of $\cA$. Quotienting this algebra by its ideal generated by the radical 
of $\cA$, one obtains a non-degenerate Clifford algebra $\bar\cC_p$. Since $\cA$ is a $G$-module, there 
exists 
an action of the little group $G_p$ on $\bar\cC_p$; let 
us denote this action by $a\mapsto g\cdot a$ for $a\in\bar\cC_p$ and $g\in G_p$. To obtain a representation 
of the full supergroup (\ref{supergroup}), one must find an irreducible representation $\tau$ of 
$\bar\cC_p$ and a representation $\cR_0$ of $G_p$ in the same space, that is compatible with $\tau$ in the 
sense that
\be
\tau[g\cdot a]
=
\cR_0[g]\cdot\tau[a]\cdot(\cR_0[g])^{-1}.
\label{compat}
\ee
For finite-dimensional groups, the pair $(\tau,\cR_0)$ turns out to be unique up to multiplication of $\cR_0$ 
by a character of $G_p$ (and possibly up to parity-reversal). Given such a pair, we call it the {\it 
fundamental representation} of the 
supersymmetric little group.
\end{enumerate}

The Clifford algebra (\ref{cliff}) leads to a replacement of the irreducible, ``spin'' representations of the 
little group, by generally {\it reducible} representations $\cR_0\otimes\cR$. This is the multiplet 
structure of supersymmetry: the restriction of an irreducible unitary representation of a supergroup to its 
bosonic subgroup is generally reducible, and the various irreducible components account for the combination 
of 
spins that gives a {\sc susy} multiplet. In the Poincar\'e group, an irreducible multiplet contains finitely 
many spins; by contrast, we will see below that a super-BMS$_3$ multiplet contains infinitely many spins. 
Apart from this difference, the structure of induced representations of super semi-direct products is 
essentially the same as in the bosonic case: they consist of wavefunctions on an orbit, taking their values 
in the space of the representation $\cR_0\otimes\cR$. \label{ROR} In particular, formula (\ref{Frob}) for the 
character remains valid, up to the replacement of $\cR$ by $\cR_0\otimes\cR$.

\subsubsection*{Supersymmetric BMS$_3$ groups}

Before turning to super BMS$_3$, recall first that the $\cN=1$ super Virasoro algebra is built by adding to 
$\text{Vect}(S^1)$ an odd subalgebra $\cF_{-1/2}(S^1)$ of $(-1/2)$-densities on the circle 
\cite{Dai,guieu2007algebre}. This produces a Lie superalgebra, isomorphic to 
$\text{Vect}(S^1)\oplus\cF_{-1/2}(S^1)$ as a vector space, which we will write as $\text{sVect}(S^1)$. Its 
elements are pairs $(X,S)$, where $X=X(\varphi)\partial/\partial\varphi$ and 
$S=S(\varphi)(d\varphi)^{-1/2}$, and the super Lie bracket is 
defined as
\be
\label{3.28}
\left[
(X,S),(Y,T)
\right]
\equiv
\left(
[X,Y]+S\otimes T,\phi_XT-\phi_YS
\right).
\ee
Here $[X,Y]$ is the standard Lie bracket of vector fields and $\phi$ denotes the natural action of 
vector fields on $\cF_{-1/2}(S^1)$, so that $\phi_XT$ is the $(-1/2)$-density whose component is
\be
XT'-\frac{1}{2}X'T.
\ee
Upon expanding the functions $X(\varphi)$ and $S(\varphi)$ in Fourier modes, one recovers the standard 
$\cN=1$ supersymmetric extension of the Witt algebra. Choosing $S(\vf)$ to be periodic or antiperiodic 
leads to the Ramond or the Neveu-Schwarz sector of the superalgebra, respectively.\\

The central extension of $\text{sVect}(S^1)$ is the super 
Virasoro algebra, $\mathfrak{svir}$. Its elements are triples $(X,S,\lambda)$ where 
$(X,S)\in\text{sVect}(S^1)$ and $\lambda\in\RR$, with a super Lie bracket defined as
\be
\left[
(X,S,\lambda),(Y,T,\mu)
\right\}
\equiv
\left(
[X,Y]+S\otimes T,\phi_XT-\phi_YS,C(X,Y)+D(S,T)
\right),
\label{superbra}
\ee
where we write
\be
C(X,Y)\equiv-\frac{1}{48\pi}\int_0^{2\pi}d\varphi\, XY'''
\quad\text{and}\quad
D(S,T)\equiv\frac{1}{12\pi}\int_0^{2\pi}d\varphi\, S'T'.
\label{cocycle}
\ee
Here $C$ is the standard Gelfand-Fuchs cocycle of the Virasoro algebra, and $C(X,Y)+D(S,T)$ 
is its supersymmetric generalisation. Again, upon expanding the functions $X$ and $S$ in Fourier modes, one 
obtains the usual commutation relations of the $\cN=1$ super Virasoro algebra, with the central charge 
$Z=(0,0,1)$.\\

We can now define the $\cN=1$ super BMS$_3$ group \cite{Barnich:2014cwa,Barnich:2015sca}: it is a super 
semi-direct 
product (\ref{supergroup}) whose even piece is the BMS$_3$ group (\ref{BMS3}), and whose odd subspace is 
$\cF_{-1/2}(S^1)$ with the bracket $\{S,T\}=S\otimes T$. In other words, the (centerless) super 
$\mathfrak{bms}_3$ algebra is a super semi-direct sum
\be \label{superbms3}
\mathfrak{sbms}_3
=
\text{Vect}(S^1)\inplus
\left(\text{Vect}(S^1)_{\text{Ab}}\oplus\cF_{-1/2}\right),
\ee
where $\text{Vect}(S^1)_{\text{Ab}}\oplus\cF_{-1/2}$ may be seen as an Abelian version of 
$\text{sVect}(S^1)$. Again, choosing periodic/antiperiodic boundary conditions for $\cF_{-1/2}$ yields the 
Ramond/Neveu-Schwarz sector of the theory (respectively). Upon including central extensions, elements of the 
(now centrally extended) super $\mathfrak{bms}_3$ algebra become $5$-tuples 
$(X,\alpha,S;\lambda,\mu)$, where $(X,\alpha,S)$ belongs to $\mathfrak{sbms}_3$ and $\lambda,\mu$ are real 
numbers, with a super Lie bracket
\bea
\left[
(X,\alpha,S;\lambda,\mu),(Y,\beta,T;\kappa,\nu)
\right\}
= 
\quad\quad\quad\quad\quad\quad\quad\quad\quad
\quad\quad\quad\quad\quad\quad\quad\quad\quad\quad\quad && \nn\\
=
\Big(
[X,Y],[X,\beta]-[Y,\alpha],\phi_XT-\phi_YS;
C(X,Y),
C(X,\beta)-C(Y,\alpha)+D(S,T)
\Big), &&
\label{superbms}
\eea
with $C(X,Y)$ and $D(S,T)$ written in (\ref{cocycle}). Upon expanding all fields in Fourier modes, one finds 
the brackets (\ref{bms3}) where the $J_m$'s and $P_m$'s are modes of $X$'s and $\alpha$'s (respectively), 
supplemented with
\begin{subequations} \label{s32.5}
\bea
[J_m,Q_r] & = & \left(\frac{m}{2}-r\right)Q_{m+r}\,,\\[1pt]
[P_m,Q_r] & = & 0\,,\\[1pt]
\{Q_r,Q_s\} & = & P_{r+s}+\frac{c_2}{6}\,r^2\delta_{r+s,0}\,,
\eea
\end{subequations}
where the supercharges $Q_r$ are the modes of $S$'s. The indices $r$, $s$ are integers/half-integers in the 
Ramond/Neveu-Schwarz sector.\\

In the gravitational context, the functions $X$ and $\alpha$ generate superrotations and supertranslations, 
while $S(\vf)$ generates local supersymmetry transformations that become global symmetries upon enforcing 
suitable boundary conditions on the fields. The surface charge associated with $(X,\alpha,S)$ then takes the 
form \cite{Barnich:2014cwa}
\be
\label{chargesusy}
Q_{(X,\alpha,S)}[j,p,\psi]
=
\frac{1}{2\pi}\int^{2\pi}_0 d\varphi \left[
X(\vf)j(\vf)+\alpha(\vf)p(\vf)+S(\vf)\psi(\vf)
\right],
\ee
where $j$ and $p$ are the angular momentum and Bondi mass aspects that we already encountered in (\ref{s25}), 
while $\psi(\vf)$ is one of the subleading components of the gravitino at null infinity. Upon using formula 
(\ref{ss25}), these charges satisfy the algebra (\ref{superbms})-(\ref{s32.5}) with $c_2=3/G$. Note that the 
gravitino naturally satisfies Neveu-Schwarz boundary conditions on the celestial circles, as it 
transforms under a projective representation of the Lorentz group.\\

The construction of the super BMS$_3$ group can be generalised in a straightforward 
way. Indeed, let $G$ be a (bosonic) group, $\mathfrak{g}$ its Lie algebra, 
$\mathfrak{sg}$ a super Lie algebra whose even subalgebra is $\mathfrak{g}$. Then one can associate with $G$ 
a (bosonic) semi-direct product $G\ltimes\mathfrak{g}$ --- the even BMS$_3$ group is of that form, with $G$ 
the Virasoro group. Now let $\mathfrak{sg}_{\text{Ab}}$ denote the ``Abelian'' super Lie algebra which is 
isomorphic to 
$\mathfrak{sg}$ as a vector space, but where all brackets involving elements of $\mathfrak{g}$ are set to 
zero. One may then define a super semi-direct product
\be
\left(
G\ltimes\mathfrak{g},
\mathfrak{g}\inplus\mathfrak{sg}_{\text{Ab}}
\right)
\ee
where we use the notation (\ref{supergroup}). This structure appears to be ubiquitous in 
three-dimensional, asymptotically flat supersymmetric higher-spin theories.\\

Unitary representations of the super BMS$_3$ group can be classified along the lines briefly explained at the 
beginning of this subsection. In the remainder of this section we describe this classification in some more 
detail and use it to evaluate characters of the centrally extended super BMS$_3$ group.

\subsubsection*{Admissible super BMS$_3$ orbits}

The unitary representations of super BMS$_3$ are classified by the same orbits as in the 
purely bosonic case. However, supermomenta that do not satisfy condition (\ref{admiss}) are forbidden, so our 
first task is to understand which 
orbits are admissible. To begin, recall that the admissibility condition (\ref{admiss}) is invariant under 
superrotations. Thus, if we consider a supermomentum orbit containing a constant $p_0$ say,
the supermomenta on the orbit will be admissible if and only if $p_0$ is. 
Including the central charge $c_2$, we ask: which pairs $(p_0,c_2)$ are such that
\be
\langle (p_0,c_2),\{S,S\}\rangle\geq 0\quad\forall\,S\in\cF_{-1/2}(S^1)\,?
\ee
Here $\langle.,.\rangle$ is the pairing between supermomenta and supertranslations, given by the terms 
pairing $p$ and $\alpha$ in the surface charges (\ref{s25}) and (\ref{chargesusy}).Using the super Lie 
bracket (\ref{superbms}), we find
\be
\langle (p_0,_2c),\{S,S\}\rangle
=
\frac{1}{2\pi}\int_0^{2\pi}d\varphi
\left(
p_0(S(\varphi))^2+\frac{c_2}{6}(S'(\varphi))^2
\right).
\label{pss}
\ee
Since the 
term involving $(S')^2$ can be made arbitrarily large while keeping $S^2$ arbitrarily small, a necessary 
condition for $(p_0,c_2)$ to be admissible is that $c_2$ be non-negative. The admissibility condition on 
$p_0$, 
on the other hand, depends on the sector under consideration:
\begin{itemize}
\item In the Ramond sector, $S(\varphi)$ is a periodic function on the circle. In particular, 
$X(\varphi)=\text{const}$ is part of the supersymmetry algebra, so for expression (\ref{pss}) to be 
non-negative for any $S$, we must impose $p_0\geq0$.
\item In the Neveu-Schwarz sector, $S(\varphi)$ is antiperiodic (i.e.~$S(\varphi+2\pi)=-S(\varphi)$) and can 
be expanded in Fourier modes as
\be
S(\varphi)=\sum_{n\,\in\,\mathbb{Z}}s_{n+1/2}e^{i(n+1/2)\varphi}.
\ee
Then expression (\ref{pss}) becomes
\be
\label{s35}
\langle (p_0,c_2),\{S,S\}\rangle
=
\sum_{n\,\in\,\mathbb{Z}}
\left[
p_0+\frac{c_2}{6}(n+1/2)^2
\right]|s_{n+1/2}|^2,
\ee
and the admissibility condition amounts to requiring all coefficients in this series to be non-negative, 
which gives
\be
p_0\geq-\frac{c_2}{24}\,.
\label{bound}
\ee
\end{itemize}
These bounds are consistent with earlier observations in three-dimensional supergravity 
\cite{Barnich:2014cwa}, according to which Minkowski space-time 
(corresponding to $p_0=-c_2/24$) realises the Neveu-Schwarz vacuum, while the Ramond vacuum is realised by 
the 
null orbifold (corresponding to $p_0=0$). Analogous results hold in AdS$_3$ \cite{Coussaert:1993jp}. More 
general admissibility conditions can be worked out for {\it non-constant} supermomenta by adapting the proof 
of the positive energy theorem in \cite{Balog:1997zz}; we will address this question elsewhere.

\subsubsection*{Super BMS$_3$ multiplets}

As explained around (\ref{cliff}), a unitary representation of super BMS$_3$ based on an orbit $\cO_p$ comes 
equipped with a 
representation $\tau$ of the Clifford algebra
\be
\label{s36}
\cC_p=T\left(\cF_{-1/2}(S^1)\right)/
\left\{
S^2-\langle(p,c_2),\{S,S\}\rangle
\right\}.
\ee
Let us build such a representation. We will work in the Neveu-Schwarz sector, and we take $p$ to be a 
constant admissible supermomentum $p_0=M-c_2/24$ with $M>0$, whose little group is $\text{U}(1)$. Then the 
bilinear form (\ref{s35}) is non-degenerate and the representation $\tau$ of the Clifford algebra (\ref{s36}) 
must be such that
\be
\label{s35.5}
\tau[Q_r]\cdot\tau[Q_s]+\tau[Q_s]\cdot\tau[Q_r]=\left(
\frac{c_2}{6}(r^2-1/4)+M\right)\delta_{r+s,0}\,,
\quad r,s\in\ZZ+1/2.
\ee
In order to make $\tau$ irreducible, we start with a highest-weight state $|0\rangle$ such that 
\mbox{$\tau[Q_r]|0\rangle=0$} for $r>0$, and generate the space of the representation by its ``descendants'' 
$\tau[Q_{-r_1}]\ldots\tau[Q_{-r_n}]|0\rangle$, $r_1>\ldots>r_n>0$. It follows from the Lie brackets (\ref{s32.5}) 
that each descendant state has spin $s+\sum_{i=1}^nr_i$, where $s$ is the spin of the state $|0\rangle$; this 
observation uniquely determines the little group representation $\cR_0$ satisfying (\ref{compat}). Thus, a 
super BMS$_3$ particle consists of infinitely many particles with spins increasing from $s$ to infinity.\\

A similar construction can be carried out for the vacuum supermomentum at $M=0$, with the subtlety that the 
Clifford algebra (\ref{s36}) (or equivalently (\ref{s35.5})) is degenerate. As explained below (\ref{cliff}), 
one needs to quotient (\ref{s36}) by the radical of the bilinear form (\ref{s35}), resulting in a 
non-degenerate Clifford algebra $\bar\cC_p$. In the case at hand this algebra is generated by supercharges 
$Q_r$ with $|r|>1$, and the representation $\tau$ must satisfy (\ref{s35.5}) with $M=0$ and $|r|,|s|>1$. The 
remainder of the construction is straightforward: starting from a state $|0\rangle$ with, say, vanishing 
spin, we generate the space of the representation by acting on it with $\tau[Q_{-r}]$'s, where $r>1$. The 
vacuum representation of super BMS$_3$ thus contains infinitely many ``spinning vacua'' with increasing spins.

\subsubsection*{Characters}

The Fock space representations just described can be used to evaluate characters. For example, in the massive 
case we find
\be
\label{s35b}
\text{tr}\left[e^{i\theta J_0}\right]
=
e^{i\theta s}\left[
1+e^{i\theta/2}+e^{3i\theta/2}+e^{2i\theta}+\cdots
\right]
=
e^{i\theta s}\prod_{n=1}^{\infty}\left(1+e^{i(n-1/2)(\theta+i\epsilon)}\right),
\ee
where we have added a small imaginary part to $\theta$ to ensure convergence of the product; the trace is 
taken in the fermionic Fock space associated with the ``highest-weight state'' $|0\rangle$. The vacuum case 
is similar, except that the product would start at $n=2$ rather than $n=1$ (and $s=0$). Note that 
(\ref{s35b}) explicitly breaks parity invariance; this can be fixed by replacing the parity-breaking Fock 
space representations $\tau$ described above by parity-invariant tensor products $\tau\otimes\bar\tau$, where 
$\bar\tau$ is the same as $\tau$ with the replacement of $Q_r$ by $Q_{-r}$. The trace of a rotation operator 
in the space of $\tau\otimes\bar\tau$ then involves the norm squared of the product appearing in 
(\ref{s35b}).\\

As explained on page \pageref{ROR}, the character of an induced representation of a super semi-direct product 
takes the same form (\ref{Frob}) as in the bosonic case, but with the character of $\cR$ replaced by that of 
a (reducible) representation $\cR_0\otimes\cR$ compatible with the Clifford algebra representation $\tau$. We 
thus find that the character of a rotation by $\theta$ (together with a Euclidean time translation by 
$\beta$), in the parity-invariant vacuum representation of the $\cN=1$, Neveu-Schwarz super BMS$_3$ group, 
reads
\bea
\chi_{\text{vac}}^{\text{super BMS}}[(\text{rot}_{\theta},i\beta)]
& = &
\chi_{\text{vac}}^{\text{BMS}}[(\text{rot}_{\theta},i\beta)]\cdot
\prod_{n=2}^{\infty}|1+e^{i(n-1/2)(\theta+i\epsilon)}|^2\nn\\
& = &
e^{\beta c_2/24}
\prod_{n=2}^{\infty}\frac{|1+e^{i(n-1/2)(\theta+i\epsilon)}|^2}{|1-e^{in(\theta+i\epsilon)}|^2}\,.
\eea
Comparing with (\ref{ss6.5q}) and (\ref{s9.5b}), we recognise the product of the partition functions of two 
massless fields with spins 2 and 3/2, that is, the one-loop partition function of $\cN=1$ supergravity in 
three-dimensional flat space.

\subsubsection*{Higher-spin supersymmetry and hypergravity}

In \cite{Fuentealba:2015jma,Fuentealba:2015wza}, the authors considered a three-dimensional hypergravity theory consisting of a metric coupled to a single field with half-integer spin $s+1/2$, with $s$ larger than one. Upon imposing suitable asymptotically flat boundary conditions, they found that the asymptotic symmetry algebra spans a superalgebra that extends the bosonic $\mathfrak{bms}_3$ algebra by generators $Q_r$ of spin $s+1/2$.  The one-loop partition function of that system is the product of the graviton partition function (see eq.~(\ref{ss6.5q}) for $s=2$) with the fermionic partition function (\ref{s9.5b}). We now show that this partition function coincides with the vacuum character of the corresponding asymptotic symmetry group (in the Neveu-Schwarz sector).\\

The irreducible, unitary representations of the asymptotic symmetry group of \cite{Fuentealba:2015wza} are classified by the same orbits and little groups as for the standard BMS$_3$ group. In particular, we can consider the orbit of a constant supermomentum $p_0=M-c_2/24$; the associated Clifford algebra representation $\tau$ mentioned below (\ref{cliff}) then satisfies a natural generalization of eq.~(\ref{s35.5}) (see eq.~(7.23) in \cite{Fuentealba:2015wza})
\be
\label{QQ}
\tau[Q_r]\tau[Q_{\ell}]+\tau[Q_{\ell}]\tau[Q_r]
=
\prod_{j=0}^{s-1}
\left(
\frac{c_2}{6}
\left(
r^2-\frac{(2j+1)^2}{4}
\right)
+M
\right)\delta_{r+\ell,0}\,,
\ee
where $r$ and $\ell$ are integers or half-integers, depending on the sector under consideration (Ramond or Neveu-Schwarz, respectively). In order for the orbit to be admissible in the sense of (\ref{admiss}), the value of $M$ must be chosen so as to ensure that all coefficients on the right-hand side of (\ref{QQ}) are non-negative. In particular, the vacuum value $M=0$ is admissible in the Neveu-Schwarz sector, in which case the anticommutators $\{\tau[Q_r],\tau[Q_{-r}]\}$ vanish for $|r|=1/2,...,s-1/2$. Thus, in the Neveu-Schwarz vacuum, the Clifford algebra (\ref{QQ}) degenerates and $\tau$ must really be seen as a representation of the non-degenerate subalgebra generated by the $Q_r$'s with $|r|\geq s$. The corresponding Fock space representation can be built as explained below (\ref{s35.5}), and the spins of the basis states in this representation are uniquely determined by the fact that the $Q_r$'s have spin $s+1/2$. The corresponding Fock space character is thus
\be
\text{tr}\left[e^{i\theta J_0}\right]
=
\prod_{n=s+1}^{\infty}
\left(1+e^{i(n-1/2)(\theta+i\epsilon)}\right),
\ee
which generalises (\ref{s35b}). The character for $\tau\otimes\bar\tau$ is the squared norm of this expression, and the resulting vacuum character of the hypersymmetric BMS$_3$ group is
\be
\chi_{\text{vac}}^{\text{hyper BMS}}
[(\text{rot}_{\theta},i\beta)]
=
e^{\beta c_2/24}\,
\frac
{\prod\limits_{n=s}^{\infty}
|1+e^{i(n+1/2)(\theta+i\epsilon)}|^2}
{\prod\limits_{m=2}^{\infty}
|1-e^{im(\theta+i\epsilon)}|^2}\,.
\ee
As announced earlier, this coincides with the one-loop partition function of asymptotically flat gravity coupled to a massless field with spin $s+1/2$.

\section{Further directions}\label{sec:future}

A first, natural extension of our work will be to compute one-loop partition functions for the missing 
particles in flat space, i.e.\ mixed-symmetry and continuous spin particles. The first case corresponds to 
representations of the little groups $SO(D-1)$ or $SO(D-2)$ with arbitrary weights, and goes beyond our 
simplifying restriction to fully symmetric fields/weights of the form $(s,0,\ldots,0)$. The second case 
corresponds instead to generic massless particles, associated with representations of the full little 
group $\text{SO}(D-2)\ltimes\RR^{D-2}$. Both setups may be physically relevant: the vast majority of string 
excitations leads to mixed-symmetry massive fields, so that any comparison between string models and  
higher-spin (gauge) theories cannot forgo a good control over mixed symmetry particles. Continuous spin 
particles are instead more elusive. For a long time, following Wigner's intuition \cite{Wigner:1939}, they 
have been considered as unphysical. Recent analyses have instead provided indications that these particles may 
even evade the standard no-go arguments against higher-spin interactions \cite{Schuster:2013pxj}. In view of 
our discussion at the beginning of sect.~\ref{sec:intro}, these representations are actually quite promising: 
in field theory they are realised by gauge theories \cite{Schuster:2014hca}; nevertheless they intrinsically 
bring in a dimensionful parameter (e.g.\ the eigenvalue of the square of the Pauli-Lubanski vector in $D=4$). 
In both cases one can easily compute the associated Poincar\'e characters with the techniques of 
sect.~\ref{sec:poincare}. Moreover, both Bose and Fermi mixed symmetry fields admit a Lagrangian description 
similar to the one we rely on in this paper (see e.g.\ \cite{Campoleoni:2008jq,Campoleoni:2009gs}), so that we 
expect to be able to smoothly extend our considerations to this class of fields. A Lagrangian description of 
continuous-spin particles has also been proposed recently \cite{Schuster:2014hca} and it will be interesting 
to test its structure by computing its one-loop partition function and comparing it with the Poincar\'e 
characters of continuous spin particles.\\

Another possible interesting application of our results will be to study carefully the flat-space limit of AdS partition functions. As we have discussed below \eqref{ss5t} and in the final remark of sect.~\ref{sec:poincare}, it is not straightforward to recover partition functions in flat space and Poincar\'e characters as limits of partition functions in AdS \cite{Gibbons:2006ij,Gupta:2012he} and characters of the conformal algebra \cite{Dolan:2005wy}. Therefore, even if considering this limit for free actions is trivial, the corresponding partition functions already give a feeling of the difficulties that become so dramatic when interactions are switched on. Clarifying how one can properly regularise the flat limit of one-loop partition functions may thus give insights on how to address this pathological limit in more general terms.\\

In three dimensions we also just started to scratch the surface of the representation theory of flat $\cW$ algebras. It will be interesting to complete the classification of coadjoint orbits and to interpret the role of classes of representations characterised by different little groups. In addition, one has to systematise the construction of the Hilbert spaces of each representation, both via the wavefunction construction that is typical of the representations of the BMS$_3$ group or via induced module constructions.

\subsection*{Acknowledgements}

We are grateful to G.~Barnich,  T.~Basile, X.~Bekaert, N.~Boulanger, D.~Grumiller, A.~Lepage-Jutier and P.~Lowdon for illuminating discussions on 
closely related topics. The research of A.C.\ and H.A.G.\ was partially supported by the ERC Advanced Grant 
``SyDuGraM'', by FNRS-Belgium (convention FRFC PDR T.1025.14 and  convention IISN 4.4514.08) and by the 
``Communaut\'e Fran\c{c}aise de Belgique" through the ARC program. The work of B.O.\ was supported by a 
doctoral fellowship of the Wiener-Anspach Foundation and by the Fund for Scientific Research-FNRS under grant 
number FC-95570. Finally, this research of M.R.\ is supported by the FWF projects P27182-N27 and the START project Y435-N16. M.R.\ is also supported by a \emph{DOC} fellowship of the Austrian Academy of Sciences, the \emph{Doktoratskolleg Particles and Interactions} (FWF project DKW1252-N27) and the FWF project I 1030-N27.


\begin{appendix}

\section{From mixed traces to bosonic characters}

\subsection{Mixed traces and symmetric polynomials}\label{AppA1}

In this part of the appendix we prove that the mixed trace (\ref{s5.5t}) of $\II_{\mu_s,\alpha_s}$ in 
$D$ 
dimensions coincides with a certain difference of complete homogeneous symmetric polynomials in the traces of $J^n$ as given by
\be
\label{s6.5}
\chi_s[n\vec\theta\,]
=
h_s(J^n)-h_{s-2}(J^n)\,,
\ee
where
\be
\label{ss6.5}
h_s(J^n)
=
\sum_{\substack{m_1,\ldots,m_s\in\,\NN\\
m_1+2m_2+\ldots+sm_s=\,s}}\!
\left[\,
\prod_{k=1}^s
\frac{\left(
{\rm Tr}[(J^n)^k]
\right)^{m_k}}{m_k!k^{m_k}}
\,\right].
\ee
By definition, the {\it complete homogeneous symmetric polynomial} of degree $s$ in $D$ complex 
variables $\lambda_1,\ldots,\lambda_D$ is
\begin{equation}
\label{s43.5}
h_s(\lambda_1,\ldots,\lambda_D)
=
\sum_{\substack{\ell_1,\ldots,\ell_D=\,0\\
\ell_1+\ldots+\ell_D=\,s}}^s
\lambda_1^{\ell_1}\lambda_2^{\ell_2}\ldots\lambda_D^{\ell_D}
=
\sum_{1\leq\ell_1\leq\ell_2\leq\ldots\leq\ell_s\leq D}
\lambda_{\ell_1}\lambda_{\ell_2}\ldots\lambda_{\ell_s}.
\end{equation}
Using the variant of Newton's identities
\be
h_s(\lambda_1,\ldots,\lambda_D)
=
\frac{1}{s}\sum_{N=1}^sh_{s-N}(\lambda_1,\ldots,\lambda_D)(\lambda_1^N+\ldots+\lambda_D^N)\,,
\ee
one can show by recursion (see e.g. \cite[p. 24f]{Macdonald1995}) that the polynomial (\ref{s43.5}) can equivalently be written as in \eqref{ss6.5}:
\be
\label{ss43.5}
h_s(\lambda_1,\ldots,\lambda_D)
=
\sum_{\substack{m_1,\ldots,m_s\in\,\NN \\ m_1+2m_2+\ldots+sm_s=\,s}}
\prod_{k=1}^s\frac{(\lambda_1^k+\ldots+\lambda_D^k)^{m_k}}{m_k!\,k^{m_k}}\,.
\ee
We will use this relation later. To prove (\ref{s6.5}), we start with the 
following:

\paragraph{Lemma.} Let $J$ be a complex $D\times D$ matrix with eigenvalues $\lambda_1,\ldots,\lambda_D$. Then,
\be
\label{eq:ProofPart1}
\left(\delta^{\mu\alpha}\right)^s\frac{1}{s!}\left(J_{\mu\alpha}\right)^s
=
h_s(\lambda_1,\lambda_2,\ldots,\lambda_D)\,,
\ee
where we use the same notation for contracting symmetrised indices as in (\ref{2.9}).

\paragraph{Proof.} The left-hand side of \eqref{eq:ProofPart1} can be seen as a trace over symmetric tensor 
powers of $J$. 
Indeed, $\delta^{\mu\alpha}J_{\mu\alpha}={\rm Tr}(J)$ is clear; as 
for $\frac{1}{2}\left(\delta^{\mu\alpha}\right)^2\left(J_{\mu\alpha}\right)^2$, one gets
\begin{equation}
\frac{1}{2}\left(\delta^{\mu\alpha}\right)^2\left(J_{\mu\alpha}\right)^2
=
\frac{1}{2}\left(
\textnormal{Tr}\!
\left(J\right)^2+\textnormal{Tr}\!\left(J^2\right)
\right)
=
\textnormal{Tr}\!
\left(
S^2\!\left(J\right)
\right)
=
\frac{1}{2}\sum_{i=1}^2
\textnormal{Tr}\!\left(J^i\right)\textnormal{Tr}\!
\left(S^{2-i}\!\left(J\right)\right),
\end{equation}
where $S^k(J)$ denotes the $k^{\textnormal{th}}$ symmetric tensor power of $J$. One can then define 
recursively
\begin{equation}
\frac{1}{s!}\left(\delta^{\mu\alpha}\right)^s\left(J_{\mu\alpha}\right)^s=\textnormal{Tr}
\left(S^s\!\left(J\right)\right)=\frac{1}{s}\sum_{i=1}^s\textnormal{Tr}\left(J^i\right)\textnormal{Tr}\left(S^{
s-i}\!\left(J\right)\right),
\end{equation}
so that $\frac{1}{s!}\left(\delta^{\mu\alpha}\right)^s\left(J_{\mu\alpha}\right)^s$ is just a 
trace in the $s^\textnormal{th}$ symmetric tensor power of the $D$-dimensional vector space $V$ on which 
$J_{\mu\alpha}$ acts as a linear operator. Now consider an eigenbasis $\{e_1,\ldots,e_D\}$ for 
$J_{\mu\alpha}$, with $J\cdot e_k=\lambda_ke_k$. Since $\frac{1}{s!}\left(J_{\mu\alpha}\right)^s$ is the 
$s^\textnormal{th}$ symmetric tensor power of $J_{\mu\alpha}$ one can construct an eigenbasis for 
$\frac{1}{s!}\left(J_{\mu\alpha}\right)^s$ by symmetrising $e_{k_1}\otimes e_{k_2}\otimes\ldots\otimes 
e_{k_D}$, with $k_1\leq k_2\leq\ldots\leq k_D$. These eigenvectors have eigenvalues 
$\lambda_{l_1}\lambda_{l_2}\ldots\lambda_{l_D}$, and 
since $\left(\delta^{\mu\alpha}\right)^s\frac{1}{s!}\left(J_{\mu\alpha}\right)^s$ 
is the trace of $\frac{1}{s!}\left(J_{\mu\alpha}\right)^s$, relation (\ref{eq:ProofPart1}) follows upon using 
the second expression of $h_s(\lambda_1,\ldots,\lambda_D)$ in (\ref{s43.5}).\hfill$\blacksquare$\\

We can now turn to the proof of (\ref{s6.5}). To this end we fix conventionally the number of terms entering the contraction of two symmetrised expressions as follows. Objects with {\it lower} indices are symmetrized with the minimum number of terms required and 
without
overall normalisation factor, while objects with {\it upper} indices are not symmetrised at all, since the symmetrisation is induced by the contraction. This specification is needed because 
terms with lower and upper indices in a contraction may have a different index structure and therefore the number of terms needed for their symmetrisation may be different. For instance
\be
\begin{split}
& A^\m B^\m C^\m D_{\m\m} E_\m \equiv A^\m B^\n C^\r \left( D_{\m\n} E_\r + D_{\n\r} E_\m + D_{\r\m} E_\n  \right) \\
& = \frac{1}{2} \left( A^\m B^\n C^\r + A^\n B^\r C^\m + A^\r B^\m C^\n +A^\m B^\r C^\n + A^\r B^\n C^\m + A^\n B^\m C^\r \right) D_{\m\n} E_\r \, .  
\end{split}
\ee\\

In order to simplify computations, we define
\begin{equation}
\label{eq:TDefinition}
T_{\mu_s,\,\alpha_s}\equiv J_{\mu\alpha}\ldots J_{\mu\alpha}\,,\qquad 
T^{[s]}\equiv 
T_{\mu_s,\,\alpha_s}\left(\delta^{\mu\alpha}\right)^s,
\end{equation}
which implies the contraction rules
\be
\label{eq:TContractionRules}
\delta^{\mu\mu}T_{\mu_s,\,\alpha_s}=2\,\delta_{\alpha\alpha}T_{\mu_{s-2},\,\alpha_{s-2}}\,,
\quad
\delta^{\alpha\alpha}T_{\mu_s,\alpha_s}=2\,\delta_{\mu\mu}T_{\mu_{s-2},\alpha_{s-2}}\,.
\ee
In terms of the tensors $T_{\mu_s,\alpha_s}$, the mixed trace (\ref{s5.5t}) can be written as 
\begin{align}
& \chi_s[n\vec\theta]
=
\frac{1}{s!}\,T_{\mu_s,\beta_s}\!
\left[\!
\left(\delta^{\mu\beta}\right)^s
\!+\sum_{m=1}^{\lfloor\frac{s}{2}\rfloor}
\frac{\left(-1\right)^ms!\left[D+2\left(s-m-2\right)\right]!!}{2^{m}m!\left(s-2m\right)!
\left[
D+2\left(s-2\right)
\right]!!}
(\delta^{\mu\mu})^m
(\delta^{\mu\beta})^{s-2m}
(\delta^{\beta\beta})^m
\right]\nn\\
& \!\stackrel{\text{\eqref{eq:TContractionRules}}}{=}\!
\frac{1}{s!}\,T^{[s]}+\sum_{m=1}^{\left[\frac{s}{2}\right]}\frac{\left(-1\right)^m\left[
D+2\left(s-m-2\right)\right]!!}{2^{m-1}m!\left(s-2m\right)!\left[D+2\left(s-2\right)\right]!!}(\delta^{
\mu\mu})^m(\delta^{\mu\beta})^{s-2m}(\delta^{\beta\beta})^{m-1}\delta_{\mu\mu}T_{
\mu_{s-2},\beta_{s-2}}.
\end{align}
To compute the trace of 
the  
$(\delta^{\mu\mu})^m(\delta^{\mu\beta})^{s-2m}(\delta^{\beta\beta})^{m-1}$ 
terms, we first change our symmetrisation from $\delta_{\mu\mu}T_{\mu_{s-2},\b_{s-2}}$ (which 
contains $\frac{s!}{2(s-2)!}$ terms) to the aforementioned product of $\delta$'s. In doing so one has to 
introduce a factor accounting for the number of terms in each structure as
\begin{subequations}
\begin{align}
\delta_{\mu\mu}T_{\mu_{s-2},\,\b_{s-2}}&
\leadsto\frac{s!}{2(s-2)!}\ \textnormal{terms},\\
\left(\delta^{\mu\mu}\right)^m\left(\delta^{\mu\beta}\right)^{s-2m}\left(\delta^{\beta\beta}\right)^{m-1}
&\leadsto\frac{s!}{2^mm!}\times\frac{(s-2)!}{2^{m-1}(m-1)!(s-2m)!}\ \textnormal{terms},
\end{align}
\end{subequations}
which implies
\begin{equation}
\chi_s[n\vec\theta]
=
\frac{1}{s!}T^{[s]}+
\sum_{m=1}^{\lfloor\frac{s}{2}\rfloor}
\frac {
\left(-1\right)^m2^{m-1}(m-1)!\left[D+2\left(s-m-2\right)\right]!!}{\left[(s-2)!\right]^2\left[D+2\left(s-2\right)\right]!!}\delta_
{\mu\mu}^m\delta_{\mu\beta}^{s-2m}\delta_{\beta\beta}^{m-1}\delta^{\mu\mu}T^{\mu_{s-2},\beta_{s-2}}.
\end{equation}
Taking into account the correct combinatorial factors one obtains
\begin{equation}
\delta_{\mu\mu}^m\delta_{\mu\beta}^{s-2m}\delta_{\beta\beta}^{m-1}\delta^{\mu\mu}=\left[D+2(s-m-1)\right]
\delta_{\mu\mu}^{m-1}\delta_{\mu\beta}^{s-2m}\delta_{\beta\beta}^{m-1}+2m\,\delta_{\mu\mu}^m\delta_{\mu\beta}^{
s-2m-2}\delta_{\beta\beta}^{m}\,,
\end{equation}
which then yields
\begin{align}
&\chi_s[n\vec\theta]=\frac{1}{s!}\,T^{[s]}+
\left(\sum_{m=1}^{\lfloor\frac{s}{2}\rfloor}
\frac{\left(-1\right)^m2^{m-1}(m-1)!\left[D+2\left(s-m-1\right)\right]!!}{\left[D+2\left(s-2\right)\right]!!}
\delta_{\mu\mu}^{m-1}\delta_{\mu\beta}^{s-2m}\delta_{\beta\beta}^{m-1}\right.\nonumber\\
\label{s47}
&\left. + \sum_{m=1}^{\lfloor\frac{s}{2}\rfloor-1}
\frac{\left(-1\right)^m2^{m}m!\left[D+2\left(s-m-2\right)\right]!!}
{\left[D+2\left(s-2\right)\right]!!}\delta_{\mu\mu}^m\delta_{\mu\beta}^{s-2m-2}\delta_{\beta\beta}^{m}
\right)\!\frac{1}{\left[(s-2)!\right]^2}\,T^{\mu_{s-2},\beta_{s-2}}.
\end{align}
Shifting $m\rightarrow m+1$ in the upper sum one can see that both sums are identical apart from the overall 
sign and the lower extremum. Thus (\ref{s47}) boils down to
\begin{equation}
\chi_s[n\vec\theta]
=
\frac{1}{s!}\,T^{[s]}-\frac{1}{\left[(s-2)!\right]^2}\,\delta_{\mu\beta}^{s-2}T^{\mu_{s-2},\beta_{s-2}}
=
\frac{1}{s!}\,T^{[s]}-\frac{1}{(s-2)!}\,T^{[s-2]}\,.
\end{equation}
Now using \eqref{eq:TDefinition} and \eqref{eq:ProofPart1} one obtains
\begin{equation}
\label{ss47}
\chi_s[n\vec\theta]
=
\frac{1}{s!}\,T^{[s]}-\frac{1}{(s-2)!}\,T^{[s-2]}
=
h_s(\lambda_1, \lambda_2,\ldots,\lambda_D)-h_{s-2}(\lambda_1, \lambda_2,\ldots,\lambda_D)\,,
\end{equation}
where $\lambda_1,\ldots,\lambda_D$ are the eigenvalues of $J^n$. (These eigenvalues are $e^{\pm in\theta_j}$ for 
$j=1,\ldots,r$, and one or two unit eigenvalues depending on whether $D$ is odd or even, respectively.) This 
leads to the desired result: since traces of powers of $J^n$ can be written as
\be
{\rm Tr}[(J^n)^k]
=
\lambda_1^k+\ldots+\lambda_D^k
\ee
in terms of the eigenvalues of $J^n$, the complete homogeneous symmetric polynomials expressed as 
(\ref{ss43.5}) exactly coincide with the combination (\ref{ss6.5}), and equation (\ref{ss47}) coincides with 
(\ref{s6.5}).

\subsection{Symmetric polynomials and $\text{SO}(D)$ characters}\label{AppA2}

In this part of the appendix we review the relation between complete homogeneous symmetric polynomials and 
characters of orthogonal groups. Most of the explicit proofs can be found in \cite{FultonHarris:1991}, 
chapter 24, to which we refer for details on the arguments exposed below. We will study separately the cases 
of odd and even $D$, and we let $r\equiv\lfloor(D-1)/2\rfloor$, with $\theta_1,\ldots,\theta_r$ the 
non-vanishing angles appearing in the rotations (\ref{2.7}).

\subsubsection*{Odd $D$}

We consider the Lie algebra $\mathfrak{so}(D)=\mathfrak{so}(2r+1)$, with rank $r$. Choosing a basis of 
$\CC^{2r+1}$ such that the Lie algebra $\mathfrak{so}(2r+1)_{\CC}$ can be written in terms of complex matrices, we may choose the Cartan subalgebra to be the subalgebra 
$\mathfrak{h}$ of $\mathfrak{so}(2r+1)_{\CC}$ consisting of diagonal matrices. As a basis of $\mathfrak{h}$ 
we choose the matrices $H_i$ whose entries all vanish, except the $(i,i)$ and $(r+i,r+i)$ entries which are 
$1$ and $-1$, respectively (with $i=1,\ldots,r$). In our convention (\ref{2.6}), the operator $H_i$ generates 
rotations in the plane $(x_i,y_i)$. Then, calling $L_i$ the elements of the dual basis (such that $\langle 
L_i,H_j\rangle=\delta_{ij}$), a dominant weight is one of the form 
$\lambda=\lambda_1L_1+\ldots\lambda_rL_r\equiv(\lambda_1,\ldots,\lambda_r)$ with 
$\lambda_1\geq\ldots\geq\lambda_r\geq0$.\\

Let $\lambda$ be a dominant weight for $\mathfrak{so}(2r+1)$. According to formula (24.28) in 
\cite{FultonHarris:1991}, the character of the irreducible representation of $\mathfrak{so}(2r+1)$ with 
highest weight $\lambda$ is
\begin{equation}
\label{eq:SO2n+1CharacterFormula}
\chi^{\text{SO}(2r+1)}_\lambda[q_1,\ldots,q_r]
=
{\rm Tr}_{\lambda}\left[q_1^{H_1}\cdots q_r^{H_r}\right]
=
\frac
{\left|
q_j^{\lambda_i+r-i+\frac{1}{2}}-q_j^{-\left(\lambda_i+r-i+\frac{1}{2}\right)}
\right|}
{\left|
q_j^{r-i+\frac{1}{2}}-q_j^{-\left(r-i+\frac{1}{2}\right)}
\right|}\,,
\end{equation}
where $q_1,\cdots q_r$ are arbitrary complex numbers\footnote{Eventually these numbers will be exponentials of 
angular potentials, so they are fugacities associated with the rotation generators 
$H_i$.}, ${\rm Tr}_{\lambda}$ denotes a trace taken in the space of the representation, and 
$\left|A_{ij}\right|$ denotes the determinant of the matrix $A$ with rows $i$ and columns $j$. This expression 
is a corollary of the Weyl character formula. Using proposition A.60 and Corollary A.46 of  
\cite{FultonHarris:1991}, it can be rewritten as
\begin{equation}
\label{eq:PropA60+CorA46}
\chi^{\text{SO}(2r+1)}_\lambda[q_1,\ldots,q_r]
=
\left|h_{\lambda_i-i+j}-h_{\lambda_i-i-j}\right|,
\end{equation}
where $h_j=h_j\left(q_1,\ldots,q_n,q_1^{-1},\ldots,q_n^{-1},1\right)$ is a complete homogeneous symmetric 
polynomial of degree $j$ in $2r+1$ variables. In particular, for a highest weight 
$\lambda_s=(s,0,\ldots,0)$ (where $s$ is a non-negative integer), the matrix appearing on the right-hand 
side of \eqref{eq:PropA60+CorA46} is upper triangular, with the entry at $i=j=1$ given by $h_s-h_{s-2}$ and 
all other entries on the main diagonal equal to one. Accordingly, the determinant in 
(\ref{eq:PropA60+CorA46}) boils down to $h_s-h_{s-2}$ in that simple case. For the rotation (\ref{2.7}) we 
may identify $q_j=e^{in\theta_j}$, and we conclude that
\be
\label{eq:SO2n+1CharacterProofFinish}
\chi^{\text{SO}(2r+1)}_{\lambda_s}[n\vec\theta]
=
\frac{\left|
\sin\left[
\left(\lambda_i+r-i+\frac{1}{2}\right)n\theta_j\right]\right|}
{\left|\sin\left[\left(r-i+\frac{1}{2}\right)n\theta_j\right]\right|}
=
h_s(J^n)-h_{s-2}(J^n)\,,
\end{equation}
where $\lambda_i=s\,\delta_{i1}$. Thus for odd $D$ the difference of symmetric polynomials in (\ref{s6.5}) is just a 
character of $\text{SO}(D)$.

\subsubsection*{Even $D$}

We now turn to the Lie algebra $\mathfrak{so}(2r+2)$, with rank $r+1$. As in the odd case we choose a basis of $\mathbb{C}^{2r+2}$ such that we can write the Lie algebra $\mathfrak{so}(2r+2)$ in terms of complex matrices and the Cartan subalgebra is generated by $r+1$ diagonal matrices $H_i$ whose entries all vanish, except 
$(H_i)_{ii}=1$ and $(H_i)_{r+1+i,r+1+i}=-1$. We call $L_i$ the elements of the dual basis, and with these 
conventions a weight $\lambda=\lambda_1L_1+\ldots+\lambda_{r+1}L_{r+1}\equiv(\lambda_1,\ldots,\lambda_{r+1})$ is 
dominant if $\lambda_1\geq\lambda_2\geq\ldots\geq\lambda_r\geq|\lambda_{r+1}|$.\\

Let $\lambda$ be a dominant weight for $\mathfrak{so}(2r+2)$. Then formula (24.40) in 
\cite{FultonHarris:1991} gives the character of the associated highest-weight representation as
\be 
\begin{split}
& \chi^{\text{SO}(2r+2)}_\lambda[q_1,\ldots,q_{r+1}]
 =
{\rm Tr}_{\lambda}\left[q_1^{H_1}\cdots q_{r+1}^{H_{r+1}}\right]\\
\label{eq:SO2nCharacterFormula}
& =
\frac
{\left|
q_j^{\lambda_i+r+1-i}+q_j^{-\left(\lambda_i+r+1-i\right)}
\right|
+
\left|
q_j^{\lambda_i+r+1-i}-q_j^{-\left(\lambda_i+r+1-i\right)}
\right|}
{\left|
q_j^{r+1-i}+q_j^{-\left(r+1-i\right)}
\right|}\,,\quad\quad\quad\quad
\end{split}
\ee
where we use the same notations as in (\ref{eq:SO2n+1CharacterFormula}), except that now $i,j=1,\ldots,r+1$. 
Note that the second term in the numerator of this expression vanishes whenever $\lambda_{r+1}=0$ (because 
the $(r+1)^{\text{th}}$ row of the matrix $q_j^{\lambda_i+r+1-i}-q_j^{-\left(\lambda_i+r+1-i\right)}$ 
vanishes). Since this is the case that we will be interested in, we may safely forget about that second term 
from now on. Alternatively, for the mixed traces (\ref{s5.5t}) that we need, we may take $q_j=e^{in\theta_j}$ 
for $j=1,\ldots,r$ and $q_{r+1}=1$ without loss of generality, so that this second term vanishes again. 
Using proposition A.64 of \cite{FultonHarris:1991}, one can then rewrite 
\eqref{eq:SO2nCharacterFormula} as
\begin{equation}
\label{eq:PropA64}
\chi^{\text{SO}(2r+2)}_\lambda[q_1,\ldots,q_r,1]
=
\left|h_{\lambda_i-i+j}-h_{\lambda_i-i-j}\right|,
\end{equation}
where $h_j=h_j\left(q_1,\ldots,q_r,1,q_1^{-1},\ldots,q_r^{-1},1\right)$. Finally, using the same arguments as for 
odd $D$, one easily verifies that the determinant on the right-hand side of \eqref{eq:PropA64} reduces once 
more to $h_s-h_{s-2}$ for a highest weight $\lambda_s=(s,0,\ldots,0)$. Writing again $q_j=e^{in\theta_j}$, 
one concludes that, for even $D$,
\begin{equation}
\label{eq:SO2nCharacterProofFinish}
\chi^{\text{SO}(2r+2)}_{\lambda_s}
\left[n\theta_1,\ldots,n\theta_r,n\theta_{r+1}=0\right]
=
\frac
{\left|
\cos\left[\left(\lambda_i+r+1-i\right)n\theta_j\right]
\right|}
{\left|
\cos\left[\left(r+1-i\right)n\theta_j\right]
\right|}
\Bigg|_{\theta_{r+1}=0}
\!=
h_s(J^n)-h_{s-2}(J^n),
\end{equation}
where $\lambda_i=s\,\delta_{i1}$. This concludes the proof of (\ref{s6b}). Note that, for {\it non-vanishing} 
$\theta_{r+1}$, the quotient of denominators in the middle of (\ref{eq:SO2nCharacterProofFinish}) is actually 
the character $\chi^{\text{SO}(2r+2)}_{\lambda_s}\left(n\theta_1,\ldots,n\theta_r,n\theta_{r+1}\right)$. This 
detail will be useful in appendix \ref{AppA3}.

\subsection{Differences of $\text{SO}(D)$ characters}\label{AppA3}

In this part of the appendix we prove the following relations between characters of orthogonal groups:
\begin{subequations}
\label{app:SO(D)DifferenceProofRelations}
\begin{align}
\chi^{\text{SO}(2r+1)}_{\lambda_s}[\vec\theta]-\chi^{\text{SO}(2r+1)}_{\lambda_{s-1}}[\vec\theta]
=
&\ \chi^{\text{SO}(2r)}_{\lambda_s}[\vec\theta]\,,
\label{app:SO(D)DifferenceProofRelations1}\\
\chi^{\text{SO}(2r)}_{\lambda_s}[\vec\theta]-\chi^{\text{SO}(2r)}_{\lambda_{s-1}}[\vec\theta]
=
&\ \sum_{k=1}^r\mathcal{A}^r_k[\vec\theta]
\chi^{\text{SO}(2r-1)}_{\lambda_{s}}[\theta_1,\ldots,\widehat{\theta_k},\ldots,\theta_r]\,.
\label{app:SO(D)DifferenceProofRelations2}
\end{align}
\end{subequations}
Here $\vec\theta=(\theta_1,..,\theta_r)$, $\lambda_s$ is the weight with components $(s,0,\ldots,0)$ in the 
basis defined above equations (\ref{eq:SO2n+1CharacterFormula}) and (\ref{eq:SO2nCharacterFormula}), and the 
hat denotes omission of an argument, while the 
coefficients $\cA^r_k$ are the quotients of determinants defined in (\ref{s6q}). Note that, when one of the 
angles $\theta_1,\ldots,\theta_r$ vanishes, say $\theta_{\ell}=0$, then $\cA^r_k=\delta_{k\ell}$ and relation 
(\ref{app:SO(D)DifferenceProofRelations2}) reduces to
\be
\label{s37.5}
\left.\chi^{\text{SO}(2r)}_{\lambda_s}[\vec\theta]\right|_{\theta_{\ell}=0}
-
\left.\chi^{\text{SO}(2r)}_{\lambda_{s-1}}[\vec\theta]\right|_{\theta_{\ell}=0}
=
\chi^{\text{SO}(2r-1)}_{\lambda_s}[\theta_1,\ldots,\widehat{\theta_{\ell}},\ldots,\theta_r]\,.
\ee

\paragraph{Proof of (\ref{app:SO(D)DifferenceProofRelations}).} We start by defining the matrices
\be
\label{A.2a}
(A^r)_{ij}=\sin\left[(r-i+\tfrac{1}{2})\theta_j\right],\qquad 
(B^r)_{ij}=\cos\left[(r-i)\theta_j\right],
\ee
so that in particular
\be
\label{s38}
\cA^r_k(\vec\theta)
=
\frac{|B^r|_{\theta_k=0}}{|B^r|}.
\ee
We will also use the shorthand notation
\be
\label{notation}
M^{r}[\theta_k]\equiv|M_{ij}(\theta_1,\ldots,\theta_{k-1},\theta_{k+1},\ldots,\theta_{r+1})|
\ee
to denote the determinant of the $r\times r$ matrix missing the angle $\theta_k$ of any of 
the matrices defined in (\ref{A.2a}). As a preliminary step towards the proof, we list the 
four following identities:
\begin{subequations}
\label{app:SO(D)DifferenceProofRelationsToProveBosonic}
\begin{align}
\frac{|A^r|}{\prod_{j=1}^r\sin\left(\theta_j/2\right)}
=&\,2^{r-1}|B^r|\,,
\label{app:SO(D)DifferenceProofRelationsToProveBosonic1}\\[3pt]
|\cos\left[(r-i)\theta_j\right]|
=&\,2^\frac{(r-1)(r-2)}{2}
\prod_{1\leq i<j\leq r}(\cos(\theta_i)-\cos(\theta_j))\,,
\label{app:SO(D)DifferenceProofRelationsToProveBosonic2}\\
\frac{|B^r|_{\theta_k=0}}{A^{r-1}[\theta_k]}
=&\,2^{r-1}(-1)^{k+1}\prod_{\substack{j=1\\j\neq k}}^r\sin\left(\theta_j/2\right),
\label{app:SO(D)DifferenceProofRelationsToProveBosonic3}\\
|B^r|
=&\,\sum_{k=1}^r|B^r|_{\theta_k=0}\,.
\label{app:SO(D)DifferenceProofRelationsToProveBosonic4}
\end{align}
\end{subequations}
Here (\ref{app:SO(D)DifferenceProofRelationsToProveBosonic1}) can be proven by induction on $r$ upon 
expanding the determinant $|A^r(\vec\theta)|$ along the first line of the matrix $A^r$. Property 
(\ref{app:SO(D)DifferenceProofRelationsToProveBosonic2}) can be shown by observing that
\begin{equation}
\cos[(r-i)\theta_j]
=
2^{r-i-1}\cos^{r-i}(\theta_j)+\sum_{k=1}^{r-i-1}c_k\cos(k\theta_j)
\end{equation}
with some irrelevant real coefficients $c_k$, and that the contribution of the second term of this expression 
to the determinant $|\cos[(r-i)\theta_j]|$ vanishes by linear dependence. Equation 
(\ref{app:SO(D)DifferenceProofRelationsToProveBosonic3}) then follows from 
(\ref{app:SO(D)DifferenceProofRelationsToProveBosonic1}) and 
(\ref{app:SO(D)DifferenceProofRelationsToProveBosonic2}), while property 
(\ref{app:SO(D)DifferenceProofRelationsToProveBosonic4}) can again be proved by induction on $r$.\\

Equipped with eqs.~(\ref{app:SO(D)DifferenceProofRelationsToProveBosonic}), we can tackle the proof of 
(\ref{app:SO(D)DifferenceProofRelations}). Equation (\ref{app:SO(D)DifferenceProofRelations1}) is easy: using 
expression (\ref{eq:SO2n+1CharacterProofFinish}) for the character $\chi_{\lambda_s}^{\text{SO}(2r+1)}$, we 
can write the difference of characters on the left-hand side of (\ref{app:SO(D)DifferenceProofRelations1}) as
\be
\chi^{\text{SO}(2r+1)}_{\lambda_s}-\chi^{\text{SO}(2r+1)}_{\lambda_{s-1}}
=
\frac
{\sum\limits_{k=1}^r(-1)^{k+1}
2\cos[(s+r-1)\theta_k]\sin\left(\theta_k/2\right)A^{r-1}[\theta_k]}
{|A^r|}\,.
\ee
Property (\ref{app:SO(D)DifferenceProofRelationsToProveBosonic1}) then allows us to reduce this expression to 
the quotient of denominators appearing in the middle of eq.~(\ref{eq:SO2nCharacterProofFinish}) (with the 
replacement of $r+1$ by $r$ and all angles non-zero), which is indeed the sought-for character 
$\chi_{\lambda_s}^{\text{SO}(2r)}[\vec\theta]$.\\

Equation (\ref{app:SO(D)DifferenceProofRelations2}) requires more work. Using once more the expression in the 
middle of (\ref{eq:SO2nCharacterProofFinish}), we first rewrite the left-hand side of 
(\ref{app:SO(D)DifferenceProofRelations2}) as
\begin{equation}
\label{app:SO(D)DifferenceProofRelationsOddDRHS}
\chi^{\text{SO}(2r)}_{\lambda_s}-\chi^{\text{SO}(2r)}_{\lambda_{s-1}}
=
\frac
{\sum\limits_{k=1}^r(-1)^{k+1}
(-2\sin[(s+r-\tfrac{3}{2})\theta_k]
\sin\left(\theta_k/2\right)B^{r-1}[\theta_k]}
{|B^r|}\,.
\end{equation}
Let us now recover this expression as a combination of characters $\text{SO}(2r-1)$: using formula 
(\ref{eq:SO2n+1CharacterProofFinish}) and the identities 
(\ref{app:SO(D)DifferenceProofRelationsToProveBosonic}), one finds
\bea
&   & \sum_{k=1}^r\chi^{\text{SO}(2r-1)}_{\lambda_s}[\theta_1,\ldots,\widehat{\theta_k},\ldots,\theta_r]
|B^r|_{\theta_k=0}\nn\\
& \stackrel{\text{(\ref{app:SO(D)DifferenceProofRelationsToProveBosonic3})}}{=} &
\sum_{k=1}^r(-1)^{k+1}2^{r-1}
\prod_{\substack{j=1\\j\neq k}}^r
\sin\left(\theta_j/2\right)
\times\nonumber\\
&   &
\times\left[
\sum_{j=1}^{k-1}(-1)^{j+1}\sin[(s+r-\tfrac{3}{2})\theta_j]A^{r-2}[\theta_j,\theta_k]
+\sum_{j=k+1}^{r}(-1)^{j}\sin[(s+r-\tfrac{3}{2})\theta_j]A^{r-2}[\theta_j,\theta_k]\right] \nonumber \\
& \stackrel{\text{(\ref{app:SO(D)DifferenceProofRelationsToProveBosonic1})}}{=} &
\sum_{k=1}^r(-1)^{k+1}2^{2r-4}\sin[(s+r-\tfrac{3}{2})\theta_k]
\sin\left(\theta_k/2\right)\times\nonumber\\
&   &
\times\left[
\sum_{j=1}^{k-1}(-1)^{j}B^{r-2}[\theta_j,\theta_k]
\prod_{\substack{i=1\\i\notin\{j,k\}}}^r\sin^2\left(\theta_i/2\right)
+\sum_{j=k+1}^r(-1)^{j+1}B^{r-2}[\theta_j,\theta_k]\prod_{\substack{i=1\\i\notin\{j,k\}}}^r
\sin^2\left(\theta_i/2\right)
\right]\nn\\
& \stackrel{\text{(\ref{app:SO(D)DifferenceProofRelationsToProveBosonic2})}}{=} &
\sum_{k=1}^r(-1)^{k+1}(-2)\sin[(s+r-\tfrac{3}{2})\theta_k]
\sin\left(\theta_k/2\right)
\left[
\sum_{j=1}^{k-1}(-1)^{j}
\left.B^{r-1}[\theta_k]\right|_{\theta_j=0}
+\sum_{j=k+1}^r\left.B^{r-1}[\theta_k]\right|_{\theta_j=0}
\right]\nn\\
& \stackrel{\text{(\ref{app:SO(D)DifferenceProofRelationsToProveBosonic4})}}{=} &
\sum_{k=1}^r(-1)^{k+1}(-2)\sin[(s+r-\tfrac{3}{2})\theta_k]
\sin\left(\theta_k/2\right)
B^{r-1}[\theta_k].
\eea
This coincides with the numerator of the right-hand side of (\ref{app:SO(D)DifferenceProofRelationsOddDRHS}), 
so identity (\ref{app:SO(D)DifferenceProofRelations2}) follows with $\cA^r_k$ given by 
(\ref{s38}).\hfill$\blacksquare$

\subsection{From $\text{SO}(D)$ to $\text{SO}(D-1)$}\label{AppA4}

In this appendix we prove relation (\ref{t6q}) between characters of $\text{SO}(D)$ and $\text{SO}(D-1)$:

\paragraph{Lemma.} 
\begin{subequations}\label{app:SO(D)SO(D-1)ProofRelations}
\begin{align}
\chi^{\text{SO}(2r+1)}_{\lambda_s}[\theta_1,\ldots,\theta_r]
=&
\sum_{j=0}^s\chi^{\text{SO}(2r)}_{\lambda_j}(\theta_1,\ldots,\theta_r),
\label{app:SO(D)SO(D-1)ProofRelations1}
\\
\chi^{\text{SO}(2r)}_{\lambda_s}[\theta_1,\ldots,\theta_r]
=&
\sum_{j=0}^s\sum_{k=1}^r
\mathcal{A}^r_k(\vec\theta)
\chi^{\text{SO}(2r-1)}_{\lambda_j}[\theta_1,\ldots,\widehat{\theta_k},\ldots,\theta_r].
\label{app:SO(D)SO(D-1)ProofRelations2}
\end{align}
\end{subequations}
Here $\lambda_j$ is the weight $(j,0,\ldots,0)$ as explained above (\ref{s6b}) or below 
(\ref{eq:PropA60+CorA46}), and $\cA^r_k(\vec\theta)$ is the quotient (\ref{s6q}) or (\ref{s38}). Since the 
proofs of these two identities are very similar, we will only display the proof of 
(\ref{app:SO(D)SO(D-1)ProofRelations1}).

\paragraph{Proof of (\ref{app:SO(D)SO(D-1)ProofRelations1}).}

Eq.~(\ref{app:SO(D)SO(D-1)ProofRelations1}) can be written as
\be
\frac{\sum\limits_{k=1}^r(-1)^{k+1}\sin[(s+r-\tfrac{1}{2})\theta_k]
A^{r-1}[\theta_k]}{|A^r|}
\stackrel{\text{(\ref{app:SO(D)DifferenceProofRelationsToProveBosonic1})\&(\ref{app:SO(D)DifferenceProofRelationsToProveBosonic3})}}{=}
\sum_{j=0}^s
\frac{\sum\limits_{k=1}^r(-1)^{k+1}\cos[(j+r-1)\theta_k]B^{r-1}[\theta_k]}{|B^r|},
\ee
where we used formulas (\ref{eq:SO2n+1CharacterProofFinish}) and (\ref{eq:SO2nCharacterProofFinish}) for the 
characters, as well as the definition (\ref{A.2a}) of $A^r$ and $B^r$. One can then use identities 
(\ref{app:SO(D)DifferenceProofRelationsToProveBosonic1}) and 
(\ref{app:SO(D)DifferenceProofRelationsToProveBosonic3}) to match the right-hand side of 
this expression with the left-hand side, proving the desired identity.\hfill$\blacksquare$

\section{From mixed traces to fermionic characters}

\subsection{Mixed traces and symmetric polynomials}\label{AppB1}

Our goal here is to prove the first equality of (\ref{s9.5}), following the same method as in appendix 
\ref{AppA1} for the bosonic case. First, using the definition (\ref{2.16}) of $U$ and the contraction rules 
(\ref{eq:TContractionRules}), one can write (\ref{s10}) as
\begin{align}
\label{app:FermionicCHSPStartingEqSimple1}
\chi_s^{(F)}[n\vec\theta]
=&
\left[
\frac{1}{s!}T^{[s]}
+
\sum_{m=1}^{\lfloor\frac{s}{2}\rfloor}
\frac
{(-1)^m[D+2(s-m-1)]!!}
{2^{m-1}m!(s-2m)![D+2(s-1)]!!}
(\delta^{\mu\mu})^m(\delta^{\mu\beta})^{s-2m}(\delta^{\beta\beta})^{m-1}\delta_{\mu\mu}
T_{\mu_{s-2},\beta_{s-2}}
\right]\!
\textnormal{Tr}[U^n]\nonumber\\
+&\sum_{m=0}^{\lfloor\frac{s-1}{2}\rfloor}
\frac{(-1)^{m+1}[D+2(s-m-2)]!!}{2^mm!(s-2m-1)![D+2(s-1)]!!}
\textnormal{Tr}
[T_{\mu_{s-1},\beta_{s-1}}
\gamma_\mu\gamma^\mu(\delta^{\mu\mu})^m(\delta^{\mu\beta})^{s-2m-1}(\delta^{\beta\beta})^m U^n],
\end{align}
where $T^{[s]}$ is the notation (\ref{eq:TDefinition}). In the first term of this expression, we shift the 
symmetrisation on the $\delta$'s i.e. we exchange upper and lower indices while taking into account the change in multiplicities of the terms involved; in all other terms, we compute one contraction with 
$\delta^{\beta\beta}$. Eq.~(\ref{app:FermionicCHSPStartingEqSimple1}) then 
simplifies to
\begin{align}
\label{app:FermionicCHSPStartingEqSimple2}
&\chi_s^{(F)}[n\vec\theta]
=
\frac{1}{s!}T^{[s]}\textnormal{Tr}[U^n]
-
\frac{1}{[(s-1)!]^2[D+2(s-1)]}\textnormal{Tr}
[T^{\mu_{s-1},\beta_{s-1}}
\gamma_\mu\gamma^\mu\delta_{\mu\beta}^{s-1}U^n]\\
&+
\sum_{m=1}^{\lfloor\frac{s}{2}\rfloor}
\left[
\frac{(-1)^m2^{m-1}(m-1)![D+2(s-m-1)]!!}{[(s-2)!]^2[D+2(s-1)]!!}
T^{\mu_{s-2},\beta_{s-2}}\delta^{\mu\mu}\delta_{\mu\mu}^m\delta_{\mu\beta}^{s-2m}\delta_{\beta\beta}^{m-1}
\textnormal{Tr}[U^n]
\right.\nonumber\\
&+
\left.
\frac{(-1)^{m+1}2{m-1}(m-1)![D+2(s-m-2)]!!}{[(s-2)!]^2[D+2(s-1)]!!}
\textnormal{Tr}
[T^{\mu_{s-2},\beta_{s-2}}
\delta^{\mu\mu}\delta_{\mu\mu}^m\delta_{\mu\beta}^{s-2m-1}
\delta_{\beta\beta}^m\gamma_\mu\gamma_\beta U^n]
\right].\nonumber
\end{align}
The $\gamma$ traces and mixed traces can now be evaluated using
\begin{subequations}
\begin{align}
\gamma^\mu\gamma_\mu\delta_{\mu\beta}^{s-1}
=&
\,[D+2(s-1)]\delta_{\mu\beta}^{s-1}-\gamma_\mu\gamma_\beta\delta_{\mu\beta}^{s-2},\\[5pt]
\delta^{\mu\mu}\delta_{\mu\mu}^m\delta_{\mu\beta}^{s-2m}\delta_{\beta\beta}^{m-1}
=&
\,[D+2(s-m-1)]\delta_{\mu\mu}^{m-1}\delta_{\mu\beta}^{s-2m}
\delta_{\beta\beta}^{m-1}\nonumber\\
&+2m\,\delta_{\mu\mu}^m\delta_{\mu\beta}^{s-2m-2}\delta_{\beta\beta}^m,\\[5pt]
\delta^{\mu\mu}\delta_{\mu\mu}^m\delta_{\mu\beta}^{s-2m-1}\delta_{\beta\beta}^{m-1}\gamma_\mu\gamma_\beta
=&
\,[D+2(s-m-1)]\delta_{\mu\mu}^{m-1}\delta_{\mu\beta}^{s-2m-1}\delta_{\beta\beta}^{m-1}
\gamma_\mu\gamma_\beta\nonumber\\
&
+4m\,\delta_{\mu\mu}^m\delta_{\mu\beta}^{s-2m-2}\delta_{\beta\beta}^{m}+2m\,\delta_{\mu\mu}^m\delta_{\mu\beta}^{s-2m-3}\delta_{\beta\beta}^{m}\gamma_\mu\gamma_\beta,
\end{align}
\end{subequations}
which yields
\begin{align}
\chi_s^{(F)}[n\vec\theta]
=&
\left[
\frac{1}{s!}T^{[s]}-\frac{1}{(s-1)!}T^{[s-1]}
\right]\textnormal{Tr}[U^n]
+
\frac{1}{[(s-1)!]^2
[D+2(s-1)]}
\textnormal{Tr}[T^{\mu_{s-1},\beta_{s-1}}\delta_{\mu\beta}^{s-2}\gamma_\mu\gamma_\beta U^n]\nonumber\\
&
-\frac{D+2(s-2)}{[(s-2)!]^2[D+2(s-1)]}T^{\mu_{s-2},\beta_{s-2}}\delta_{\mu\beta}^{s-2}\textnormal{Tr}[U^n]
\nonumber\\
&
+\frac{1}{[(s-2)!]^2[D+2(s-1)]}\textnormal{Tr}[T^{\mu_{s-2},\beta_{s-2}}\delta_{\mu\beta}^{s-3}
\gamma_\mu\gamma_\beta U^n].
\end{align}
Using (\ref{ss47}) and the definition (\ref{2.16}) of $U$, together with some careful counting, one verifies 
that this expression matches $\left[h_s(J^n)-h_{s-1}(J^n)\right]\textnormal{Tr}[U^n]$, which was to be proven.

\subsection{Symmetric polynomials and $\text{SO}(D)$ characters}\label{AppB2}

In this part of the appendix we prove the second equality in (\ref{s9.5}), following essentially the same 
steps as in appendix \ref{AppA2}. We refer again to \cite{FultonHarris:1991} for details, and we write the 
components of weights in the dual basis of the Cartan subalgebra described above (\ref{eq:SO2n+1CharacterFormula}) and  
(\ref{eq:SO2nCharacterFormula}). We will consider separately odd and even space-time dimensions.

\subsubsection*{Odd $D$}

The character of a half-spin representation of $\mathfrak{so}(2r+1)$ with a dominant highest weight
$\lambda=(\lambda_1+\tfrac{1}{2},\lambda_2+\tfrac{1}{2},\ldots,\lambda_r+\tfrac{1}{2})$ is 
\cite[p.258f]{Littlewood:1950}
\begin{equation}
\chi^{\text{SO}(2r+1)}_{\lambda}\left[\theta_1,\ldots,\theta_r\right]
=
\frac
{\left|\sin\left[\left(\lambda_i+r-i+1\right)\theta_j\right]\right|}
{\left|\sin\left[\left(r-i+\frac{1}{2}\right)\theta_j\right]\right|}
=
\left(\prod_{i=1}^r2\cos\left(\tfrac{\theta_i}{2}\right)\right)
\frac{\left|\sin\left[\left(\lambda_i+r-i+1\right)\theta_j\right]\right|}
{\left|\sin\left[\left(r-i+1\right)\theta_j\right]\right|}.
\end{equation}
Owing to expression (\ref{ss9.5}) for the trace of $U^n$, the second equality in (\ref{s9.5}) is equivalent to
\begin{equation}
\label{eq:SO2n+1FermionicCharacterProofStart}
h_s(J)-h_{s-1}(J)
=
\frac
{\left|\sin\left[\left(\lambda_i+r-i+1\right)\theta_j\right]\right|}
{\left|\sin\left[\left(r-i+1\right)\theta_j\right]\right|}
\end{equation}
for $\lambda_i=s\delta_{i1}$. To prove this, consider the difference of the bosonic character 
(\ref{eq:SO2n+1CharacterProofFinish}) and the right-hand side of 
(\ref{eq:SO2n+1FermionicCharacterProofStart}):
\be
\label{ss50b}
\frac{\left|
\sin[(\lambda_i+r-i+\tfrac{1}{2})\theta_j]
\right|}{\left|
\sin[(r-i+\tfrac{1}{2})\theta_j]
\right|}
-
\frac{\left|
\sin[(\lambda_i+r-i+1)\theta_j]
\right|}{\left|
\sin[(r-i+1)\theta_j]
\right|}\,.
\ee
Introducing the notation
\be
\label{A.2b}
(\cA^r)_{ij}=2\sin\left[(r-i+1)\theta_j\right],
\quad 
(\cB^r)_{ij}=2\cos\left[(r-i+\tfrac{1}{2})\theta_j\right]
\ee
and in terms of (\ref{A.2a}), this difference can be written as
\be
\label{s50b}
\frac
{\sum_{k=1}^r(-1)^{k+1}\sin[(s+r-\tfrac{1}{2})\theta_k]A^{r-1}[\theta_k]}
{|A^r|}
-
\frac
{\sum_{k=1}^r(-1)^{k+1}2\sin[(s+r)\theta_k]\cA^{r-1}[\theta_k]}
{|\cA^r|}
\ee
upon expanding the determinants along the first row. Now it turns out that\footnote{See 
e.g.~\cite[p.259]{Littlewood:1950}.}
\be
2^r\left|A^r\right|
\prod_{i=1}^r2\cos\left(\theta_i/2\right)
=
|\cA^r|,
\quad
2^{r-1}
\left|B^r\right|
\prod_{i=1}^r2\cos\left(\theta_i/2\right)
=
|\cB^r|,
\ee
and plugging this property in (\ref{s50b}) one sees that (\ref{ss50b}) is just $h_{s-1}(J)-h_{s-2}(J)$. Since the 
first term of (\ref{ss50b}) equals $h_s(J)-h_{s-2}(J)$ by 
virtue of (\ref{eq:SO2n+1CharacterProofFinish}), this proves (\ref{eq:SO2n+1FermionicCharacterProofStart}).

\subsubsection*{Even $D$}

The character of an irreducible representation of $\mathfrak{so}(2r+2)$ with (dominant) highest-weight 
$\lambda=(\lambda_1+1/2,\ldots,\lambda_{r+1}+1/2)$ can be written as \cite[p.258-259]{Littlewood:1950}

\begin{equation}
\chi^{\text{SO}(2r+2)}_{\lambda}
\left[\theta_1,\ldots,\theta_r\right]
=
\frac{\left|\cos\left[\left(\lambda_i+r-i+\tfrac{3}{2}\right)\theta_j\right]\right|}
{\left|\cos\left[\left(r-i+1\right)\theta_j\right]\right|}
=
\prod_{i=1}^{r+1}2\cos\left(\tfrac{\theta_i}{2}\right)
\frac{\left|\cos\left[\left(\lambda_i+r-i+\tfrac{3}{2}\right)\theta_j\right]\right|}
{\left|\cos\left[\left(r-i+\tfrac{3}{2}\right)\theta_j\right]\right|},
\end{equation}
where we are including the possibility of a non-zero angles $\theta_{r+1}$ (while in (\ref{s9.5}) we take 
$\theta_{r+1}=0$). Taking into account (\ref{ss9.5}), proving the second equality in (\ref{s9.5}) amounts to 
showing that
\begin{equation}
\label{eq:SO2nFermionicCharacterProofStart}
h_s(J)-h_{s-1}(J)
=
\frac{\left|\cos\left[\left(\lambda_i+r-i+\tfrac{3}{2}\right)\theta_j\right]\right|}
{\left|\cos\left[\left(r-i+\tfrac{3}{2}\right)\theta_j\right]\right|}
\Bigg|_{\theta_{r+1}=0}
\end{equation}
for $\lambda_i=s\delta_{i1}$. To prove this we proceed as in the odd-dimensional case: the difference of the 
bosonic character (\ref{eq:SO2nCharacterProofFinish}) and the right-hand side of 
(\ref{eq:SO2nFermionicCharacterProofStart}),
\be
\left.\frac
{\left|
\cos[(\lambda_i+r-i+1)\theta_j]
\right|}
{\left|
\cos[(r-i+1)\theta_j]
\right|}
\right|_{\theta_{r+1}=0}
-
\left.\frac
{\left|
\cos[(\lambda_i+r-i+\tfrac{3}{2})\theta_j]
\right|}
{\left|
\cos[(r-i+\tfrac{3}{2})\theta_j]
\right|}
\right|_{\theta_{r+1}=0},
\ee
can be written as
\begin{equation}
\left[
\frac{\sum\limits_{k=1}^{r+1}(-1)^{k+1}\cos[(s+r)\theta_k]B^r[\theta_k]}
{|B^{r+1}|}
-
\frac{\sum\limits_{k=1}^{r+1}(-1)^{k+1}2\cos[(s+r+\tfrac{1}{2})\theta_k]\cB^r[\theta_k]}
{|\cB^{r+1}|}
\right]_{\theta_{r+1}=0}
\end{equation}
upon expanding the determinants along the first row and using the notation (\ref{A.2a})-(\ref{A.2b}). One can 
then verify that this reduces to $h_{s-1}(J)-h_{s-2}(J)$ by the same argument as in the odd-dimensional case. By 
virtue of the second equality in (\ref{eq:SO2nCharacterProofFinish}), this proves 
(\ref{eq:SO2nFermionicCharacterProofStart}).

\end{appendix}


\end{document}